\documentclass[10pt]{wlscirep}
\usepackage[utf8]{inputenc}
\usepackage[T1]{fontenc}
\usepackage{pmboxdraw} \usepackage[scaled=0.8]{beramono}
\usepackage{multicol}
\usepackage{placeins}
\usepackage{changepage}
\usepackage{amsmath}
\usepackage{tabularx}
\usepackage{algorithm} \usepackage{algpseudocode}
\usepackage{comment}

\usepackage{color}
\usepackage{fancyvrb}

\DefineVerbatimEnvironment{Highlighting}{Verbatim}{commandchars=\\\{\}}
\usepackage{framed}
\definecolor{shadecolor}{RGB}{241,243,245}
\newenvironment{Shaded}{\begin{snugshade}}{\end{snugshade}}

\newcommand{\AttributeTok}[1]{\textcolor[rgb]{0.40,0.45,0.13}{#1}}

\newcommand{\CommentTok}[1]{\textcolor[rgb]{0.37,0.37,0.37}{#1}}

\newcommand{\ConstantTok}[1]{\textcolor[rgb]{0.56,0.35,0.01}{#1}}

\newcommand{\DecValTok}[1]{\textcolor[rgb]{0.68,0.00,0.00}{#1}}
\newcommand{\DocumentationTok}[1]{\textcolor[rgb]{0.37,0.37,0.37}{\textit{#1}}}

\newcommand{\FunctionTok}[1]{\textcolor[rgb]{0.28,0.35,0.67}{#1}}

\newcommand{\NormalTok}[1]{\textcolor[rgb]{0.00,0.23,0.31}{#1}}

\newcommand{\OtherTok}[1]{\textcolor[rgb]{0.00,0.23,0.31}{#1}}

\newcommand{\SpecialCharTok}[1]{\textcolor[rgb]{0.37,0.37,0.37}{#1}}

\newcommand{\StringTok}[1]{\textcolor[rgb]{0.13,0.47,0.30}{#1}}

\usepackage{longtable,booktabs,array}
\usepackage{calc} \usepackage{etoolbox}
\makeatletter
\patchcmd\longtable{\par}{\if@noskipsec\mbox{}\fi\par}{}{}
\makeatother
\makeatother
\makeatletter
\@ifpackageloaded{tcolorbox}{}{\usepackage[skins,breakable]{tcolorbox}}
\makeatother
\makeatletter
\@ifundefined{shadecolor}{\definecolor{shadecolor}{rgb}{.97, .97, .97}}
\makeatother

\title{Cast vote records: A database of ballots from the 2020 U.S. Election}

\author[1,$\dag$,*]{Shiro Kuriwaki}
\author[2,$\dag$]{Mason Reece}
\author[2]{Samuel Baltz}
\author[3]{Aleksandra Conevska}
\author[2]{Joseph R. Loffredo}
\author[3]{Can Mutlu}
\author[1]{Taran Samarth}
\author[2]{Kevin E. Acevedo Jetter}
\author[2]{Zachary Djanogly Garai}
\author[2]{Kate Murray}
\author[4]{Shigeo Hirano}
\author[5]{Jeffrey B. Lewis}
\author[3]{James M. Snyder, Jr.}
\author[2]{Charles H. Stewart, III}
\affil[1]{Yale University, Institution for Social and Policy Studies, New Haven, CT, 06511, USA}
\affil[2]{Massachusetts Institute of Technology, Department of Political Science, Cambridge, MA, 02139, USA}
\affil[3]{Harvard University, Department of Government, Cambridge, MA, 02138, USA}
\affil[4]{Columbia University, Department of Political Science, New York, NY, 10027, USA}
\affil[5]{University of California Los Angeles, Department of Political Science, Los Angeles, CA, 90095, USA}

\affil[*]{Corresponding author: shiro.kuriwaki@yale.edu}

\affil[$\dag$]{These authors contributed equally to this work.}

\begin{abstract}
Ballots are the core records of elections.
Electronic records of actual ballots cast (\emph{cast vote records}) are available to the public in some jurisdictions.
However, they have been released in a variety of formats and have not been independently evaluated. Here we introduce a database of cast vote records from the 2020 U.S. general election. We downloaded publicly available unstandardized cast vote records, standardized them into a multi-state database, and extensively compared their totals to certified election results. Our release includes vote records for President, Governor, U.S. Senate and House, and state upper and lower chambers -- covering 42.7 million voters in 20 states who voted for more than 2,204
 candidates.
This database serves as an uniquely granular administrative dataset for studying voting behavior and election administration.
Using this data, we show that in battleground states, 1.9 percent of solid Republicans (as defined by their congressional and state legislative voting) in our database split their ticket for Joe Biden, while 1.2 percent of solid Democrats split their ticket for Donald Trump.
\end{abstract}

\begin{document}
\maketitle
\flushbottom
\thispagestyle{empty}

\section*{Background \& Summary}

Ballots are the core records of elections.
While the \emph{total counts} of ballots for individual candidates are reported regularly\cite{road97, ansolabehere14, precincts22, VEST:2022}, \emph{individual} ballots are rarely made available.

\begin{figure}[t]
\caption{\textbf{Cast Vote Record Example}. An example of an actual ballot image (left) and the authors' representation of the associated cast vote record (right) in Wisconsin. Blank marks are recorded as an undervote.}
\label{fig:cvr-example}
\includegraphics[width=\linewidth]{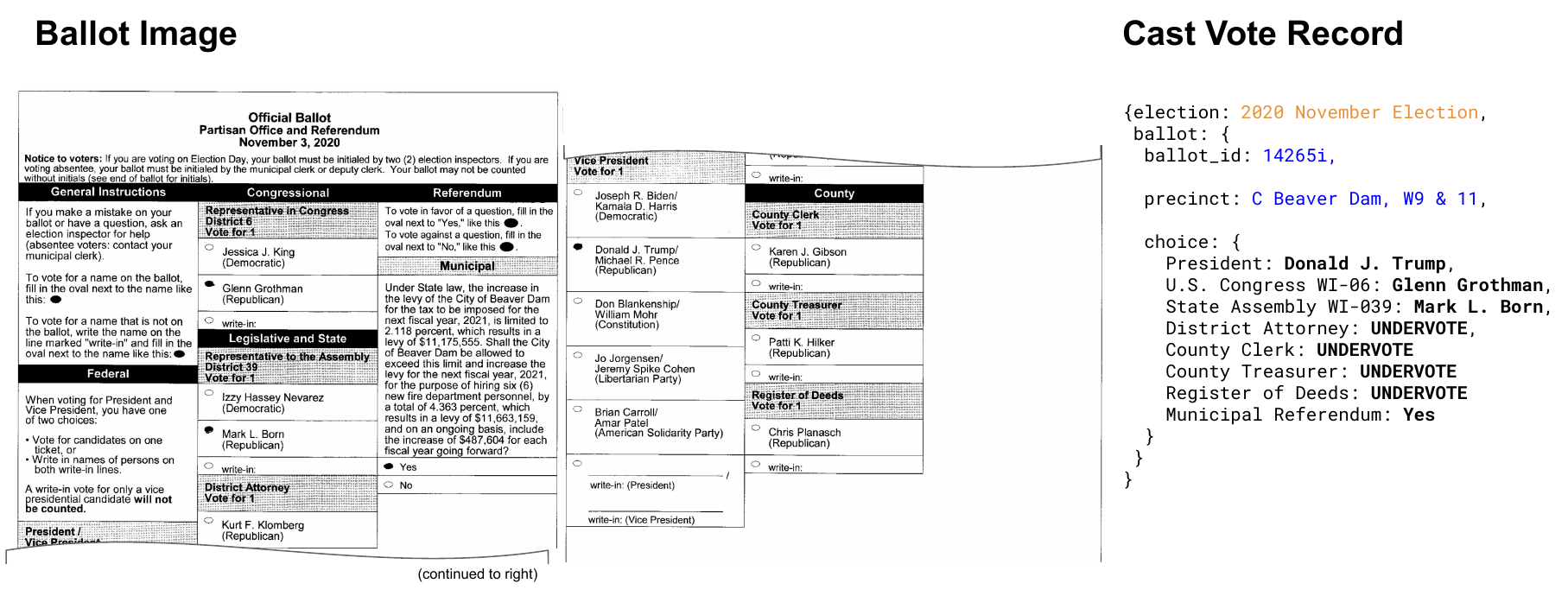}
\end{figure}

In paper-based elections, three types of records at the individual ballot level have been available to some researchers and litigators.
The first is the actual, marked paper ballot.
The second is the electronic scan of the paper ballot, often referred to as a ballot image.
The third is an electronic record of the machine's interpretation of that scanned record, called \emph{cast vote records} (CVRs, \autoref{fig:cvr-example}).  The National Institute of Standards and Technology describes CVRs as ``an electronic record of a voter’s selections, with usually one CVR created per sheet (page) of a ballot.''\cite{wack2019cast}. 
CVRs are not the ultimate basis of an election, but because ``election results are [often] produced by tabulating the collection of CVRs''\cite{wack2019cast}, they should directly reproduce vote totals produced by ballot tabulators.
Among the three types of records, CVRs lends itself best for analysis.

In this study, we introduce a dataset of CVRs representing 42.7 million
 \unskip \ voters. The dataset is available on the \emph{Dataverse} at \url{https://doi.org/10.7910/DVN/PQQ3KV}\cite{dataverse}.
Unlike certified election results --- typically made available by state election officials --- CVRs are rarely centralized at the state level. Instead, CVRs are often produced as a byproduct of the tabulation process conducted at the sub-state level and retained by local election officials.

Following the November 3, 2020 U.S. general election, local election officials saw a surge in requests for cast vote records by anonymous constituents.
Some states saw a four to five-fold increase in records requests between 2020 and 2022\cite{green2023foia,leingang2022}.
Owing to the work of election administrators in responding to these requests, and to O'Donnell for posting those unprocessed records online\cite{odonnell}, researchers now have access to an unprecedented quantity of cast vote records.  O'Donnell and his collaborators originally collected the CVRs in order to investigate the validity of the 2020 presidential election\cite{bloombergnews2022} (to evaluate these claims, we refer readers to work by other researchers\cite{grimmer2024evaluating}).
We standardized the publicly available CVRs into a single database, and extensively compared them to official election results.

Our final dataset has two main audiences.
The first audience are those in political science, economics, sociology, and other related fields who study electoral behavior. CVR data will allow researchers to study a wide variety of electoral phenomena.
In particular, it will allow researchers to measure important aspects of voting behavior much more accurately than is possible using aggregate election returns or surveys. The CVRs are at the individual level and record voters’ actual choices across multiple offices. For example, the CVRs include choices for state legislative races. These are rarely included in surveys because researchers typically focus on top-of-the-ticket races.  Moreover, survey estimates of vote choices in down-ballot elections can result in especially large measurement errors because respondents are less likely to correctly recall who they supported in down-ballot elections. Such estimates may have especially large sampling errors too, because only a few respondents in a typical nationally representative sample will have participated in any particular down-ballot contest such as those for state legislative seats.

One important electoral phenomenon is split-ticket voting, that is, voting for candidates with different partisan affiliations across offices. For example, aggregate-level election data can tell us how many votes Donald Trump received in a state and how many voted for each of the Republican candidates for offices further down the ballot, but it cannot tell us how many voters who voted for Donald Trump for President \emph{also} voted for Republicans all on the down-ballot offices\cite{burden2009split,rogers2016national, malzhanhall}.
CVRs allow researchers to measure this type of behavior precisely, and at different levels of government, e.g., national level, state level, and across levels\cite{reece2024hidden, ConevskaMPSA2024, kuriwaki_JMP}.
Researchers can also use CVRs to count the number of voters who vote for Democratic candidates and also vote for the progressive position in referendums\cite{dubin1992patterns, Gerber2004, besley2008issue}.

The data include geographic identifiers: counties and often precincts. One application possible with CVRs is to measure which state legislative or congressional districts have an especially high share of split-ticket voters. To the degree that these voters can be viewed as “swing” or “persuadable” voters, while straight-ticket voters can be viewed as “core” or “loyal” partisan voters, this measure could be used to test theories of resource allocation and campaign strategies\cite{stromberg2008electoral,larcinese2013testing,hill2017changing}.  

Researchers can of course merge the CVR data with other information about the candidates and contests, such as incumbency status, patterns of campaign spending, and candidate attributes, such as race, ethnicity, gender, ideology, and/or experience\cite{dowling2024SC}. With this extra data in hand, researchers can begin to study a wide variety of phenomena, including which voters split their tickets and in which ways as a function of the available choices. They can also study the degree to which split-ticket voting favors incumbents or the candidate who has a fundraising advantage\cite{kuriwaki_JMP}. Researchers can also merge the CVR data with precinct-level demographic and socio-economic information to explore relationships between split-ticket voting and these types of variables. For example, does ticket splitting vary with the types and amounts of media (e.g., local newspapers) available in an area\cite{DeLuca2022}?   

The CVR data could be used to study other electoral phenomena as well. For example, CVRs for ranked choice elections allows researchers to analyze voter's preferences across all candidates\cite{alvarez2018low,atsusaka2024analyzing}. Or, consider roll-off, where voters cast ballots but choose to abstain in particular races (also called undervoting). Using the CVRs, researchers can measure this behavior accurately. They can then investigate the types of races where this behavior is more prevalent, and the types of voters (e.g., straight-ticket Democrats, straight-ticket Republicans, or ticket-splitters) that are more likely to roll-off, as well as the interaction between contest and voter types.
One final example is voting for candidates of minor parties, such as the Libertarian Party and the Green Party. Since the CVR database has millions of records, it contains many thousands of records of individuals who supported minor parties that received a small fraction of the vote\cite{Herron2007}. The typical survey is less useful for studying this type of behavior due to sample size.

The second audience are those in election law, forensics, and administration, who seek to study the integrity of the electoral process\cite{adler13, bernhard2017public}.
Cast vote records are of interest to scholars of attitudes towards election integrity because voters' mistrust of the vote counting process, when it exists, often revolves around the ballots\cite{gerber2013there, atkeson23, jaffe2023effect}.
Ballot-level data have been used in studies of election administration to help explain seemingly surprising election results\cite{wand2001butterfly, Bafumi2012}.
Our dataset can also contribute to debates around the tradeoff between transparency and privacy\cite{biggers2023can, gerber2013perceptions, williams2024votes,kuriwaki23}.

Finally, our data has implications that go beyond the particular states in this release, or even the 2020 election in particular.
The November 2020 election was a turning point in U.S. politics, where the administration of elections became an overtly partisan issue.
The conduct and administration of any other future elections now risks being politicized.
Self-enforcing the legitimacy of elections requires election administrators, data scientists, and social scientists to work together upon a common understanding of election technology.
We hope that our data release serves as a standard for future work in this area of growing interest.

\section*{Methods}

Out of 3143 counties in the U.S., \texttt{votedatabase.com} releases data from 464 individual counties and 3 statewide data sets (Alaska, Delaware, and Rhode Island). As described by O'Donnell\cite{odonnell}, this website provides access to the raw electronic file containing CVRs that were acquired by numerous citizens via open records requests. Of the 467 files, we validate and release 362 counties in 20 states. We additionally modify precinct information from 0.003\% of the data for privacy protection. Privacy and our technical validation are discussed in more detail below.

We approached the raw data cautiously.  CVRs may not include all of the ballots cast in an election for a number of reasons.
Even within a single county, it is possible that some valid ballots are cast and tabulated in a way that creates CVRs while others are not. Sometimes ballots that are held aside for manual adjudication, such as provisional or damaged ballots, are not scanned through tabulators, and therefore a cast vote record is not created for such ballots.  We can increase our confidence that CVR files posted are complete, uncorrupted, and genuine by comparing vote tallies produced using the downloaded CVR data to the official reports of vote totals from the same jurisdictions.  Note that our goal is \emph{not} an audit of the election.  CVRs are neither sufficient nor necessary to rule out election malfeasance.  While CVRs are convenient representations of the paper ballot, they are simply not intended to be the official results of an election.  
On the other hand, even if the cast vote record were complete, ballots could have been tampered prior to tabulation. 

We start with data that O'Donnell and his collaborators obtained and subsequently uploaded to \texttt{votedatabase.com}\cite{odonnell}. O'Donnell reports that citizens requested CVRs according to their state's open record law guidelines, noting that all records were ``obtained through these valid public records requests''\cite{odonnell}. He reports that requests were sent to ``nearly all counties in all states,''  
but 23 states responded that they did not have any records relevant to the request that could be provided under the state's open records law (Alabama, Connecticut, Hawaii, Indiana, Kansas, Louisiana, Maine, Massachusetts, Mississippi, Missouri, Montana, Nebraska, New Hampshire, New York, North Carolina, North Dakota, Oklahoma, South Carolina, South Dakota, Utah, Virginia, Washington, and Wyoming).  Indeed, the availability of CVRs is limited by state law and executive order, with states including North Carolina, South Carolina, and New York shielding the CVR from open records requests\cite{kuriwaki23}.  In all, the CVRs in the data presented here originate from counties in 27 states and D.C. that were collected and made available by O'Donnell\cite{odonnell}.  The database contains data from swing states like Wisconsin, Michigan, and Georgia, as well as solidly Democratic states, including New Jersey, and solidly Republican states, including Texas.

Our standardization and validation proceeded in five steps.
First, we downloaded the data files and standardized them so that their values were comparable across different voting machines and jurisdictions.
While the website does provide some normalized versions of the raw data, we have chosen not to rely on these and instead independently process the raw data.
We only considered files that were clearly CVRs in machine-readable form covering multiple offices and more than a handful of precincts.
Standardization here entails that we identify the party affiliation of each candidate, code invalid votes consistently across jurisdictions, and standardize the formatting of candidate names.
In the initial phase of the study, two groups conducted these pursuits independently, without awareness of each other's work.
This gave us nearly independent measures of inter-coder reliability and reduced the possibility that a single coding error propagated to all counties.

About 15 percent of the counties had CVRs in non-rectangular formats such as JSON or XML. The remaining files were tabular files such as CSV or Microsoft Excel.
Formats differed by the vendor of the machine (the main three being Dominion, Hart Intercivic, and ES\&S), and the make of each machine.  For information about the particular machine used in each county, see \url{https://verifiedvoting.org/verifier}.
We parsed these data with our own R and Python scripts, eventually normalizing all data into a long format where one row represents a single vote choice by a voter. This article releases the full codebase we developed.

Second, we assigned a cast vote record identifier to each voter, within a county or jurisdiction. Most times, this number indicates an individual voter -- one anonymous voter gets one identifier for all their choices.
In about 30 counties with ballots spanning multiple double-sided pages, each page was separated before it was scanned.  (Note that from the administrator's perspective, once the ballot itself is deemed valid, the identity of the voter is irrelevant in the counting process.)
In about 20 of these counties, we used metadata to link pages into a single ID (see Supplementary Information A).
In the remaining 10, the records were irreversibly separated. This includes  several counties in California (Los Angeles, San Francisco, San Bernardino, Ventura Counties).
However, in the remaining counties, there is a one-to-one correspondence between the ID we assign and a single voter.
Even in many counties with long ballots such as Maricopa, Arizona (with around 60 contests per ballot), the ballot record for each individual voter is preserved.

Third, we extensively checked the CVRs against other official sources of data, mainly the MIT Election Data and Science Lab (MEDSL) 2020 precinct-level returns\cite{medsl2020bystate,precincts22}. This was an ideal dataset to use as validation because it is at the precinct-level, it has standardized its formatting across states, it includes district-level as well as statewide contests, and features extensive documentation.
We limited our attention to six offices: U.S. President, Governor, U.S. Senate, U.S. House, State Senate, and State House.
The CVRs include votes for many other offices, including local administrative offices, school boards, and referendums.
Other work analyzes these offices\cite{ConevskaMPSA2024, reece2024hidden} but we exclude them in this dataset because we lack fully standardized official data to extensively validate against.
After attempting to find the best matching official result for all counties, we only release counties with candidate-level discrepancies of 1 percent or less at the candidate-county level.
The section ``Technical Validation'' discusses the process in more depth.

Fifth, we extracted precinct information in the cast vote records and standardized them to match the MIT Election Data and Science Lab database of standardized precinct names.
Precinct identifiers were often either names (e.g., "City of Madison Ward 1") or numeric codes (e.g., Precinct 35-001).
These classifications are known to vary widely across jurisdictions and machines, with no national standard.
We used fuzzy string matching and triangulation of vote counts to link the precincts, which we detail in Supplementary Information B.
There are 312
 \unskip \ counties which we matched to the precinct level database.

\section*{Data Records}

Our dataset includes 166,470,734
 \unskip \ rows,  with each row indicating a choice of a voter for a given contest.
Our data is deposited in Dataverse at \url{https://doi.org/10.7910/DVN/PQQ3KV}\cite{dataverse}.

\paragraph{Variables}

\begin{table}[tbp]
    \centering
    \small
    \begin{tabular}{ll}
    \toprule
        Name & Description \\
    \midrule
         \texttt{state} & The name of the state that the ballot is from \\
         \texttt{county\_name} & The name of the county the ballot is from \\
         \texttt{cvr\_id} & A unique ID given to each ballot within a state-county \\
         \texttt{precinct\_medsl} & One field that indicates the precinct that can be matched to Baltz \textit{et al.} (2022).\cite{precincts22}\\
         \texttt{precinct\_cvr} & Concatenated precinct values as recorded in the raw data, for each \texttt{precinct\_medsl}.\\
         \texttt{office} & The name of the office the voter is choosing a candidate in \\
         \texttt{district} & The district of the office the voter is choosing from \\
         \texttt{candidate} & The name of the candidate the voter has selected in the office-district \\
         \texttt{party} & A simplified name of the party of the candidate on the ballot \\
         \texttt{party\_detailed} & The full name of the party of the candidate on the ballot \\
         \texttt{magnitude} & The number of candidates a voter could have chosen in this particular contest \\
    \bottomrule
    \end{tabular}
    \caption{\textbf{Dataset Variables}}
    \label{tab:codebook}
\end{table} 
\autoref{tab:codebook} describes the variables in the data. Voters are uniquely identified by the state, county name, and \texttt{cvr\_id} assigned by ourselves, with the exception of unmerged fragmented ballots described in the methods section.
Contests are uniquely identified by the state, office, and district. 
The names and values for our variables generally follow the naming conventions of the MIT Election Data and Science Lab\cite{precincts22}.

The \texttt{candidate} value is the name of the candidate that ballot was cast for, or it can be an undervote (\texttt{UNDERVOTE}), overvote (\texttt{OVERVOTE}), or write-in (\texttt{WRITEIN}).
Undervotes refer to a blank choice for that contest or mark that the tabulator could not interpret, and overvotes refer to a voter marking more candidates than they could vote for in a given contest.
Even though both types of votes are invalid, undervoting in particular is seen as a form of contest-specific abstention and is of interest to election scholars.

In all offices but the State House in Arizona and West Virginia, the contests represented here are single-choice elections (a \texttt{magnitude} of one).  
Arizona voters can cast up to 2 valid votes to elect their Representatives for the state's lower chamber, with 2 winners per district. Until 2020,  West Virginia voters in some districts could cast up to 2 or 3 valid votes for their lower chamber.

Third-party candidates and write-in candidates rarely win themselves, but the ideological orientation of voters who vote for them is of interest to researchers\cite{Herron2007}.
When a candidate is listed on the ballot with a registered party, they are listed with their party affiliation in the \texttt{party\_detailed} variable (such as Libertarian, Green, or ``No Party Affiliation'' in Florida).
The \texttt{party\_detailed} variable is set to missing for undervotes and overvotes.
Jurisdictions vary in whether write-ins are reported separately or grouped together simply as write-in votes. 
Ballot access for third parties also varies by state.
When the candidate is not listed on the ballot with a party, we record them as write-in candidates with no party affiliation (For more details, see Supplementary Information C).
 
Our dataset also records the precinct of the cast vote record, as discussed in the Methods section.
We provide two variables for precincts (\autoref{tab:codebook}).
\texttt{precinct\_medsl} is our matched precinct name, formatted to correspond exactly with those described by Baltz \textit{et al.}\cite{precincts22}.
\texttt{precinct\_cvr} is a concatenation of the original precinct or precinct portion name as it appears in the cast vote record by the pipe character \texttt{|}.  
The concatenation occurs for every set of precincts for a given \texttt{precinct\_medsl}.
For example, if precinct \texttt{001} as defined in the MEDSL precinct data\cite{precincts22} contains two precinct portions \texttt{001A} and \texttt{001B}, and the cast vote records record the precinct portion, we give all voters in precinct \texttt{001} the value of \texttt{001A | 001B}.

\paragraph{Privacy Considerations}
For certain vote records, we further aggregate the values of \texttt{precinct\_cvr} and \texttt{precinct\_medsel} by concatenating additional precinct portions and by concatenating two more MEDSL precincts.  In those records,  the \texttt{precinct\_cvr} and \texttt{precinct\_medsl} fields indicate that the vote was cast in one of the listed precincts, but not which one. We apply this additional aggregation to avoid facilitating the linkage of any particular voter to a vote that they cast. 

To understand how highly-disaggregated election results can reveal the vote choices of particular voters, suppose there is a precinct in which only 3 voters shared a particular ballot style. Here we refer to the ballot style as the subset of contests that the voter could vote on that are among the six we report. \emph{Who} voted in U.S. elections is public via lists of voter rolls produced by election officials. Those voter rolls contain sufficient geographic information to infer the identity of each of these voters from their precinct and ballot style. If all of the voters in that precinct who used that ballot style supported the same candidate, the vote choice for those 3 voters would be revealed by the CVR data to anyone who has access to an accurate voter list\cite{kuriwaki23}. While the release of CVRs might seem to increase opportunities for unraveling the secret ballot, nearly all of the information that can be used to reveal vote choices from the CVR data is already contained in the official precinct aggregates that counties commonly made public \cite{kuriwaki23}.

If we reported the \texttt{precinct\_cvr} and \texttt{precinct\_medsl} fields without additional aggregation, as many as 5,391 votes choices out of 166.7 million vote choices (0.003\%) contained in the database are considered \emph{revealed} as defined by Kuriwaki, Lewis, and Morse\cite{kuriwaki23}, i.e., can theoretically be linked to the voters who cast them, using the precinct and ballot style information contained in the database. We note that each of these 5,391 linkages is only theoretical because of practical data limitations that can sometimes make it impossible to infer the ballot style from the CVR or to construct a complete and accurate enumeration of every voter in a precinct who used a particular ballot style.

To address this potential linkage of votes to voters, we first compiled a list of precinct and ballot style combinations in which every voter supported the same candidate in at least one contest.  In each of those instances, knowing who voted in the precinct using that ballot style is sufficient to learn one or more of the vote choices made by each of those voters.  To avoid this revelation, we then pooled those revealed voters with voters from another precinct in the same county who used the same ballot style.  In particular, each revealed precinct-ballot style voter combination was concatenated with the smallest precinct-ballot style voter combination in the same county that shared the same ballot style.  For example, suppose the voters using ballot style \texttt{A} in precinct \texttt{001} had their votes revealed because all of them supported Donald Trump for President, and suppose precinct \texttt{002} had the smallest number of voters in the county among all non-revealed precinct-ballot style combinations using ballot style \texttt{A}. 
Then, we report the precinct value for voters using style \texttt{A} in both precincts as \texttt{001 | 002}.  Note that for other ballot styles used in precincts \texttt{001} and \texttt{002} (if any), there may be no vote revelations.  For this reason, the database may contain records in which precincts \texttt{001} and \texttt{002} are concatenated (those associated with voters using 
ballot style \texttt{A}) and others in which they appear alone (those associated with voters using other ballot styles).

\paragraph{Geographic Coverage}

\begin{figure}[p]
\centering
\includegraphics[width=0.9\linewidth]{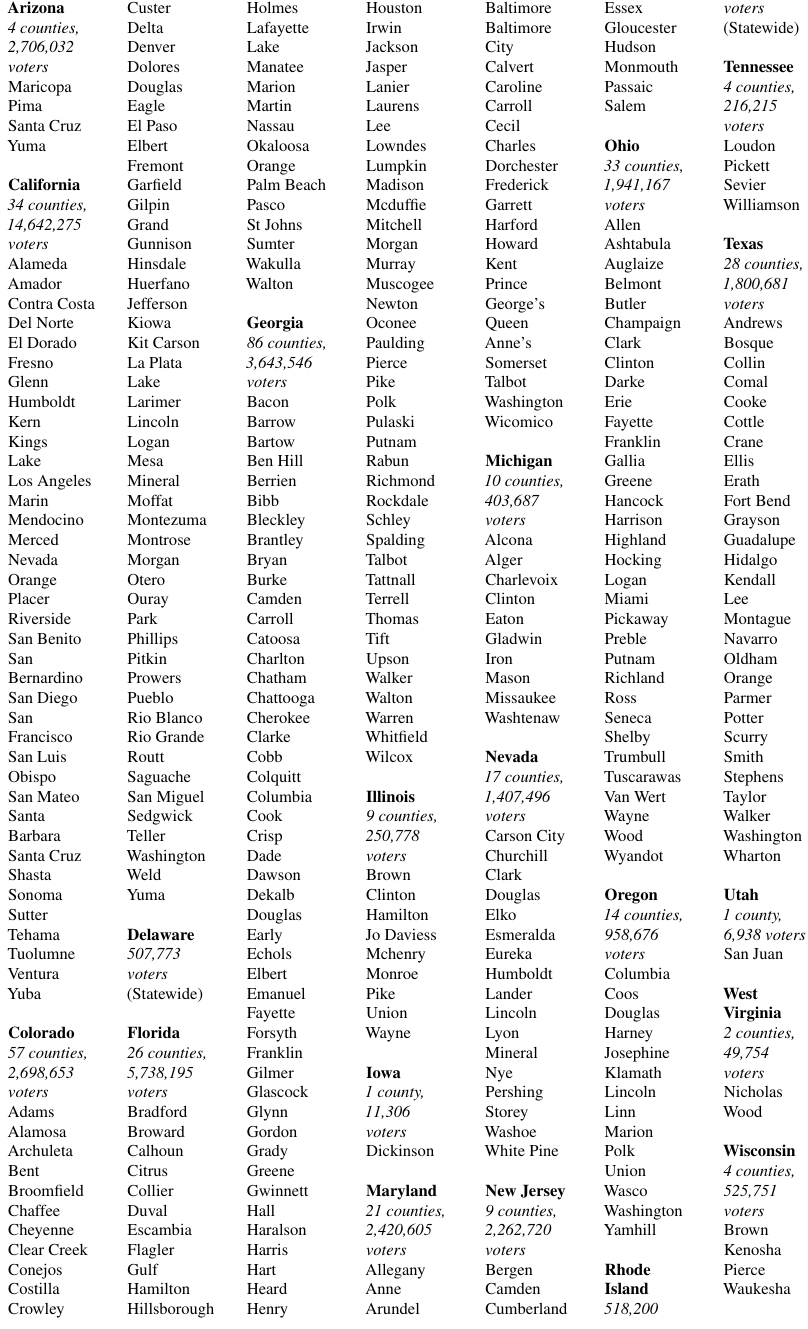}
    \caption{\textbf{List of Counties Included}. For a spreadsheet version of this list, see file \texttt{county\_info.xlsx} in our Dataverse repository. For a map of this list, see Supplementary Information D.}
    \label{tab:counties-list}
\end{figure}

A total of 20 states are covered by our release. \autoref{tab:counties-list} lists all counties in Version 2 of our released dataset.

We release all counties in the states of Nevada, Rhode Island, and Delaware after our procedure, while in other states, only a portion of the state's counties is released. 
\autoref{tab:tilt} compares the characteristics of our collected samples with the entire state.
A convenient cross-state metric to indicate the representativeness of our release is the percentage of voters that vote for the Democratic presidential candidate, Joseph R. Biden.
\autoref{tab:tilt} compares the percentage of Biden voters as a share of Biden and Trump voters in the dataset and the state(s) as a whole. 
Overall, 56.6 percent of our collection's Presidential voters are Biden voters (as a percentage of the two-party vote), while his two-party vote share in all 50 states was 52.2 percent\cite{FEC2020}.

\begin{table}[tb]
\centering
\fontsize{9.8pt}{11.7pt}\selectfont
\begin{tabular*}{0.85\linewidth}{@{\extracolsep{\fill}}lrrrrrrr}
\toprule
 & \multicolumn{3}{c}{Percent Biden} &  & \multicolumn{3}{c}{Biden, Trump, and Jorgensen Voters} \\ 
\cmidrule(lr){2-4} \cmidrule(lr){6-8}
State & CVR & Pop. & Diff. &  & CVR & Pop. & \% \\ 
\midrule\addlinespace[2.5pt]
Arizona & 52.7 & 50.2 & +3 &  & 2,677,978 & 3,385,294 & 79 \\ 
California & 65.3 & 64.9 & 0 &  & 14,245,273 & 17,305,067 & 82 \\ 
Colorado & 54.5 & 56.9 & -2 &  & 2,640,797 & 3,221,419 & 82 \\ 
Delaware & 59.6 & 59.6 & 0 &  & 500,227 & 501,871 & 100 \\ 
Florida & 50.4 & 48.3 & +2 &  & 5,694,485 & 11,036,075 & 52 \\ 
Georgia & 48.8 & 50.1 & -1 &  & 3,623,410 & 4,998,482 & 72 \\ 
Illinois & 41.4 & 58.7 & -17 &  & 247,034 & 5,985,350 & 4 \\ 
Iowa & 33.0 & 45.8 & -13 &  & 11,191 & 1,676,370 & 1 \\ 
Maryland & 64.9 & 67.0 & -2 &  & 2,382,347 & 2,994,925 & 80 \\ 
Michigan & 59.3 & 51.4 & +8 &  & 400,193 & 5,511,059 & 7 \\ 
Nevada & 51.2 & 51.2 & 0 &  & 1,387,901 & 1,387,957 & 100 \\ 
New Jersey & 61.5 & 58.1 & +3 &  & 2,209,816 & 4,523,390 & 49 \\ 
Ohio & 43.7 & 45.9 & -2 &  & 1,917,001 & 5,901,568 & 32 \\ 
Oregon & 51.4 & 58.3 & -7 &  & 938,860 & 2,340,413 & 40 \\ 
Rhode Island & 60.6 & 60.6 & 0 &  & 512,238 & 512,461 & 100 \\ 
Tennessee & 31.5 & 38.2 & -7 &  & 212,808 & 2,998,363 & 7 \\ 
Texas & 42.5 & 47.2 & -5 &  & 1,787,191 & 11,272,753 & 16 \\ 
Utah & 46.8 & 39.3 & +8 &  & 6,747 & 1,463,868 & <1 \\ 
West Virginia & 27.0 & 30.2 & -3 &  & 49,261 & 792,053 & 6 \\ 
Wisconsin & 43.0 & 50.3 & -7 &  & 520,911 & 3,279,229 & 16 \\ \midrule
All 20 States & 56.6 & 53.4 & +3 &  & 41,965,669 & 91,087,967 & 46 \\ 
All 50 States & — & 52.2 & — &  & — & 155,291,327 & — \\ 
\bottomrule
\end{tabular*}

 \medskip
\caption{\textbf{Comparison of Data Coverage to Entire State}. \emph{ The first set of columns compares the two-party Biden vote share in our data (CVR) and the entire state (Pop.).} \emph{The second set of columns shows the total number of Biden, Trump, and Jorgensen votes.}}
\label{tab:tilt}
\end{table}

\autoref{tab:census} compares the set of counties based on demographics.
We take demographic data from the 2020 decennial Census at the county level and compare our counties with the entire United States. 
The Table reports the average and median value of each demographic variable, as well as the overall (population-weighted) value and the standard deviation. 
The population in our set of counties does not differ from the average U.S. county in terms of  age or homeownership, but our counties are less White (52.8\% vs. 61.2\% nationwide), more Hispanic, and more urban.

\begin{table}[tb]
\centering
\fontsize{9.8pt}{11.7pt}\selectfont
\begin{tabular*}{\linewidth}{lrrrrrrrr}
\toprule
 & \multicolumn{2}{c}{Overall} & \multicolumn{2}{c}{Average} & \multicolumn{2}{c}{Median} & \multicolumn{2}{c}{Std. Dev.} \\ 
\cmidrule(lr){2-3} \cmidrule(lr){4-5} \cmidrule(lr){6-7} \cmidrule(lr){8-9}
 & CVR & Nation & CVR & Nation & CVR & Nation & CVR & Nation \\ 
\midrule\addlinespace[2.5pt]
Percent White & 52.8 & 61.2 & 71.3 & 75.3 & 74.4 & 82.1 & 17.0 & 19.9 \\ 
Percent Black & 11.7 & 12.3 & 10.0 & 8.7 & 3.8 & 2.2 & 13.4 & 14.0 \\ 
Percent Hispanic & 27.5 & 19.5 & 14.8 & 11.9 & 9.6 & 4.8 & 14.7 & 19.2 \\ 
Percent Under 18 & 22.0 & 22.0 & 21.7 & 21.9 & 22.0 & 21.9 & 3.4 & 3.4 \\ 
Percent Over 65 & 16.3 & 16.9 & 19.4 & 20.2 & 18.5 & 19.9 & 5.3 & 4.7 \\ 
Percent Urban & 89.4 & 80.1 & 51.3 & 37.1 & 56.4 & 34.8 & 34.3 & 34.2 \\ 
Percent Homeowning & 22.2 & 24.2 & 27.0 & 28.5 & 27.0 & 29.0 & 4.7 & 4.4 \\ 
\bottomrule
\end{tabular*}

\medskip
\caption{\textbf{Characteristics of Counties Used}. \textit{Comparison of the counties in our sample (CVR) with all counties in the United States (Nation). All statistics are computed using data from the 2020 Decennial Census at the county level.}}
\label{tab:census}
\end{table}

\section*{Technical Validation}

We extensively analyzed each county's files and compared them with official results, as described in the previous section.

\paragraph{Validation} In general terms, we define a discrepancy as any difference between the summed total of CVRs cast for candidates in the CVRs we processed from a county and the certified results published by that county. More precisely,  we compute the discrepancy in a county $k$ by the percentage
\begin{align}
\text{discrepancy}_{k} = \text{max}_{j \in \mathbb{I}_{k}}\left\{\left |\frac{N^{c}_{j} - N^{v}_{j}}{N^{v}_{j}}\right | \right\},
\end{align}
where $N^c$ is the CVR vote count, $N^v$ is the official vote count, $j$ indexes candidates in a county, $|\cdot|$ is the absolute value function, and $\mathbb{I}_k$ is the set of all Republican, Democratic, and Libertarian candidates in the six offices in county $k$.
We limited our validation to these three parties because affiliations for other choices can be reported in different ways by the county or voting machine, leading to mismatches even when the count is correct. 
For example, all our released counties (except for Santa Cruz, Arizona and Cumberland, New Jersey) include undervotes, but some counties do not report the number of undervotes in their official reports so we cannot verify all these numbers.
We relied on MEDSL's precinct data\cite{precincts22} as our official vote count, and complemented it with official statements of the vote from counties directly where necessary.

\begin{figure}[tbp]
\caption{\textbf{Discrepancy Rates}. \emph{(a) number of counties by discrepancy. (b) distribution of discrepancies at the candidate level, limited to the neighborhood of 1 percent. (c) relationship between differences in total votes.}}
\label{fig:error-rates}
\includegraphics[width=\linewidth]{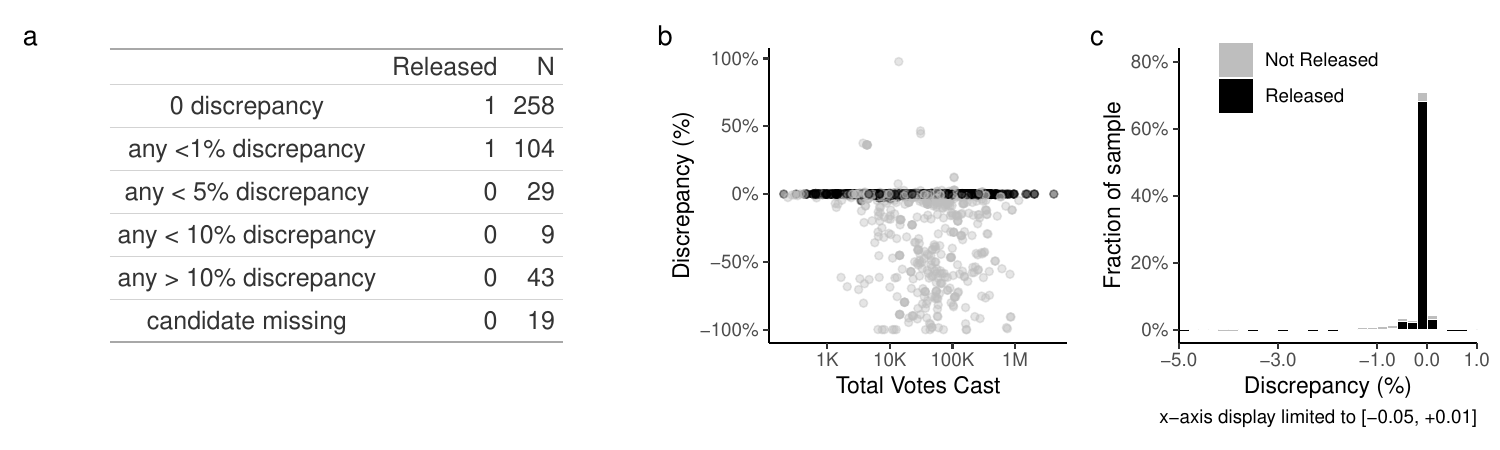}
\end{figure}

We then release counties where the discrepancy is 1 percent or less.
\autoref{fig:error-rates} shows the distribution of discrepancies.
In panel \ref{fig:error-rates}a, we grouped the discrepancy rates in six categories, showing that 258 counties had no discrepancy across all six offices, and 104 had a candidate with non-zero discrepancies within 1 percent.
Panel \ref{fig:error-rates}b shows the candidate-level discrepancies, focusing on the range of -5 to 1 percent. We see that almost all the discrepancies are within 0.2 percent (and most are exactly zero).
Some data points with 0 discrepancy are nevertheless not released and shown in light gray. This is because those counties have zero discrepancies in some offices but a discrepancy larger than 1 percent in other offices.
Panel \ref{fig:error-rates}c shows the correlates of discrepancy by population. The discrepancies tend to be dispersed across large and small jurisdictions.

\begin{figure}[tbp]
\caption{\textbf{Precinct Level Validation}.
\emph{Comparisons of vote totals for each Presidential candidate at the precinct level, with those from a precinct-level database\cite{precincts22} on the x-axis and those from our cast vote record database (with approximately matched precinct) on the y-axis. Axes are shown on a square root scale.}
}
\label{fig:precinct-val}
\centering
\includegraphics[width=0.8\linewidth]{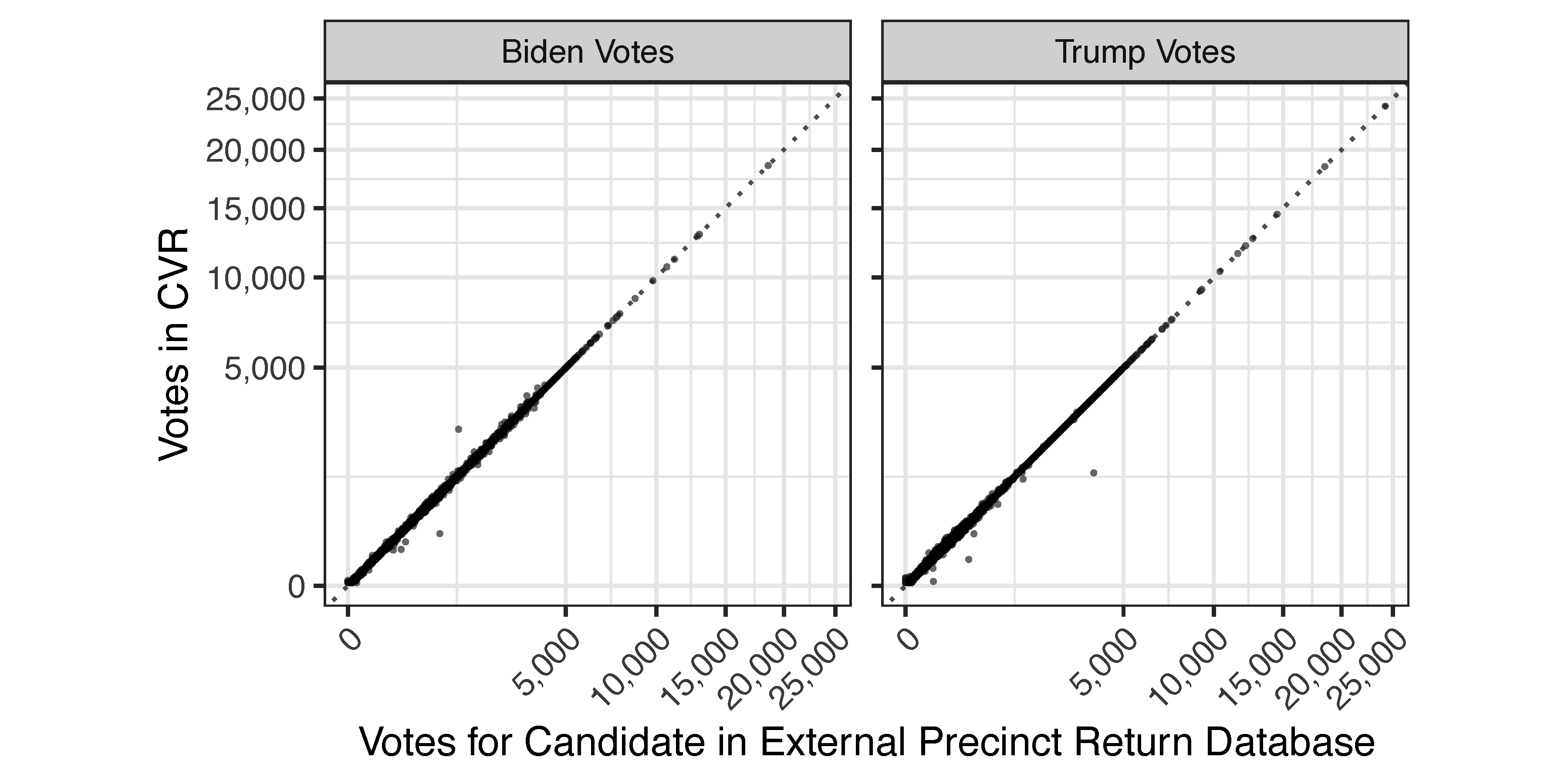}
\end{figure}

In addition to computing discrepancies at the county-level, we also conducted a precinct-level validation.
Our matching procedure described previously produced 23,139 precincts in 312 counties that could be matched to MEDSL's standardized precinct names\cite{precincts22}.
\autoref{fig:precinct-val} shows the alignment between the total votes for Presidential candidates in the CVR, per precinct, and the independently reported\cite{precincts22} votes in the precinct we matched to.
The match is not perfect due to the inclusion of counties with up to 1 percent discrepancies, and possible incorrect matches. Nevertheless, 20,139 precincts (out of 23,139 assigned) matched exactly, and 21,773 matched within 3 votes.

\FloatBarrier

\paragraph{Reasons for Discrepancy}

In the rare cases of discrepancies, we conducted an extensive, county-by-county investigation to reconcile the numbers, using state and county certification reports. 
Our public Github repository's \emph{Issue} feature documents our diagnosis for each problematic county that contains up-to-date diagnoses (\url{https://github.com/kuriwaki/cvr_harvard-mit_scripts/issues}). 
We found that the discrepancies tended to fall in one one of several categories. 

First, in some counties, entire precincts were missing from the cast vote record data.  This can happen if a county chooses to hand-count a batch of its ballots, or if the precinct's ballots are processed differently.

Second, in some counties the cast vote record data did not include votes cast by certain methods (vote by mail, in-person, provisional).  For example, in Santa Rosa County, Florida and Cuyahoga County, Ohio, it appears that the mailed-in votes have not been included in the cast vote record export uploaded to O'Donnell's\cite{odonnell} website.  We verified this by comparing our counts with the county's published vote totals broken out by vote method.  This may occur if mailed and in-person votes are handled by a different tabulator.

Another type of known discrepancy is due to redactions done by counties to protect the integrity of the secret ballot, as we previewed in the Data Description section.
Thirteen counties in Colorado and California provided CVRs where the vote choices of ballots in small ballot styles were removed. 
This practice is one way in which jurisdictions have tried to balance their dual roles of transparency and privacy.
However, perfect redaction is known to be difficult because counties also need to report the total number of votes cast with perfect fidelity, and the redacted information can be backed out by triangulating the total election results.
We do not attempt to unredact the CVRs.

There were many other counties with smaller discrepancies that could not be resolved with publicly available data.
We believe one possible explanation for these minute discrepancies is the designation of provisional and disputed votes and ballot that were hand-tabulated.
Our cast vote record may either exclude or include votes that were disputed but were later counted towards a major party candidate in the certified tally.  For example, Dodge County, Wisconsin's website cautions that 
\begin{quote}
``Cast Vote Record (CVR) Reports are unofficial results from election night. These are the results the voting equipment tabulated on Election Day. The final, official canvass results posted on the  Wisconsin Elections Commission’s website for any state/federal races also include counted provisional ballots and other small adjustments. These adjustments are not tallied by, or in, the voting equipment,  [but] rather through the County Board of Canvass process. The Cast Vote Record (CVR) Reports contain all data fields available in the ES\&S Election Software. Also, please note that if a Municipal Clerk has accidentally corrupted their election data after printing their results tapes and electronically transferring the results into the County for a specific election, that data will not be able to be archived and therefore, would have no ballots to be read and included in the CVR Report.'' (\url{https://bit.ly/46eISNX}, accessed July 1, 2024)
\end{quote}

While the dispute resolution process is publicly recorded on video in many counties, we cannot link each resolved ballot to a record in our database.

\begin{figure}[tbp]
\caption{\textbf{Reasons for Discrepancies}. An example from precinct-level matches in Dane County, Wisconsin}
\label{fig:dane}
\centering
\includegraphics[width=0.8\linewidth]{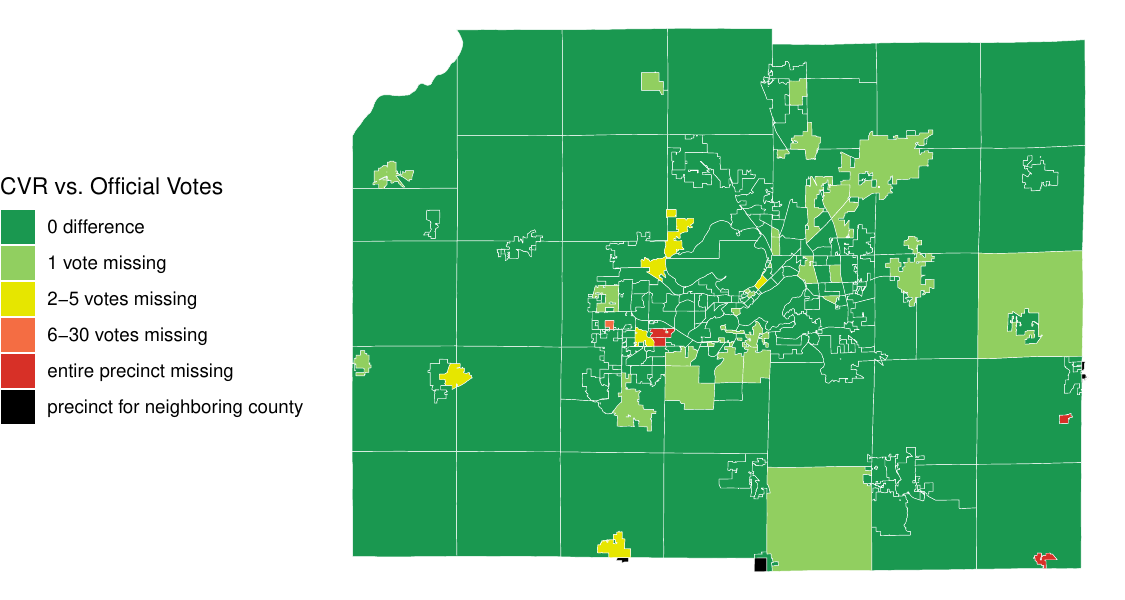}
\end{figure}

The case of Dane County, Wisconsin illustrates how several of these discrepancies can manifest.
\autoref{fig:dane} shows a precinct map of the county, which includes the City of Madison, using shapefiles\cite{VEST:2022} and each precinct colored by type of discrepancy compared to official reported results.
Two precincts were missing from the Dane CVR because they were held by the neighboring county.
Wisconsin cities and villages, which conduct their own election tabulation, may be located near a county border and straddle two counties.
In three other instances, we found that a cast vote record from a town actually belonged to the jurisdiction of a different county.
We learned that in two other precincts in Dane County, scanning machine failures prompted a hand-count of the votes, with no ballot image or cast vote record present.
In 25 other precincts, the cast vote record was only 1 vote short of the reported election result.
We contacted the Dane County clerk to resolve these issues, and obtained a letter written by the clerk's office in 2021 that provided these explanations for the discrepancies\cite{mcdonnell2021}.

We chose to limit our inquiry to county clerks to a minimum, given that many are already at or above capacity in their day-to-day duties\cite{bpcturnover}.
We believe it is unlikely that the original collectors of the data tampered with these remaining cases before posting because discrepancies are small and inconsequential for the apparent winner of the contest.

\section*{Usage Notes}
In the remaining section, we illustrate how users can read in the data and analyze it.
Although we use R as the running example, Python or any database-friendly programming language can read the dataset.
We conclude with an example that studies the party loyalty of Republican and Democratic voters in their choice for President.

\paragraph{Reading in the Data}

We store our dataset in a parquet format. Parquet is a modern file
storage format optimized for querying large datasets. It is partitioned
by grouping variables, and it is columnar (so that users do not need to
read in an entire row to extract a value from one column). Our dataset
is prohibitively large to read and write in a plain-text format (20 Gb),
but is compact and easy to read from in parquet (700 Mb). In R, we use
the \texttt{arrow} package to query parquet files. For more information
on how to read and write parquet files in R, see
\url{https://r4ds.hadley.nz/arrow}. Parquet is also designed for usage
in Python (\url{https://arrow.apache.org/docs/python/parquet.html}) and
several other programming languages.

The following command opens the dataset.

\begin{Shaded}
\begin{Highlighting}[]
\FunctionTok{library}\NormalTok{(tidyverse)}
\FunctionTok{library}\NormalTok{(arrow)}

\NormalTok{ds }\OtherTok{\textless{}{-}} \FunctionTok{open\_dataset}\NormalTok{(}\StringTok{"cvrs"}\NormalTok{)}
\end{Highlighting}
\end{Shaded}

\noindent Here, \texttt{"cvrs"} indicates the path to the top-level
folder containing the parquet files downloaded from Dataverse. Our data
is organized by county, nested within states. After unzipping the zip
file or by previewing on Dataverse, we see that \texttt{cvrs} has the
following structure:

\begin{Shaded}
\begin{Highlighting}[]
\NormalTok{├── state=ARIZONA}
\NormalTok{│   ├── county\_name=MARICOPA}
\NormalTok{│   │   └── part{-}0.parquet}
\NormalTok{│   ├── county\_name=PIMA}
\NormalTok{│   │   └── part{-}0.parquet}
\NormalTok{│   ├── county\_name=SANTA\%20CRUZ}
\NormalTok{│   │   └── part{-}0.parquet}
\NormalTok{│   └── county\_name=YUMA}
\NormalTok{│       └── part{-}0.parquet}
\NormalTok{...}
\NormalTok{── state=UTAH}
\NormalTok{│   └── county\_name=SAN\%20JUAN}
\NormalTok{│       └── part{-}0.parquet}
\NormalTok{└── state=WISCONSIN}
\NormalTok{    ├── county\_name=BROWN}
\NormalTok{    │   └── part{-}0.parquet}
\NormalTok{    ...}
\NormalTok{    └── county\_name=WAUKESHA}
\NormalTok{        └── part{-}0.parquet}
\end{Highlighting}
\end{Shaded}

Because parquet is columnar, users will find it much faster to produce
summary statistics of the data. Even though the code below counts some
166 million rows, it performs the count in one second on a personal
laptop. 

\begin{Shaded}
\begin{Highlighting}[]
\NormalTok{ds }\SpecialCharTok{|\textgreater{}} \FunctionTok{count}\NormalTok{(office) }\SpecialCharTok{|\textgreater{}} \FunctionTok{collect}\NormalTok{()}
\end{Highlighting}
\end{Shaded}

\begin{verbatim}
# A tibble: 6 x 2
  office              n
  <chr>           <int>
1 US PRESIDENT 42710448
2 US SENATE    19667514
3 US HOUSE     42613049
4 STATE SENATE 22146498
5 STATE HOUSE  38770601
6 GOVERNOR       562624
\end{verbatim}

To perform this count, we used \texttt{count()} from \texttt{dplyr} (a package loaded via \texttt{tidyverse}),
which totals the number of occurrences of each unique value in our
\texttt{office} variable. We make use of R's pipe operator,
\texttt{\textbar{}\textgreater{}}, to pass our data objects forward onto
subsequent operations we want to perform. We then used the \texttt{collect()} command from \texttt{arrow}
to extract the summary. All transformations before \texttt{count()} are
\emph{lazily-loaded}, meaning that they are not executed until needed.
The arrow program combines the transformations internally in a way that
avoids duplicative operations.

\paragraph{Application: Biden and Trump's Party Loyalty}

Here we ask whether partisans --- defined by their
votes for Congress and state legislature --- vote for their party's
presidential candidate. Donald Trump was a polarizing candidate.
Election observers have wondered if Trump drew less support from
Republican-leaning voters compared to the support his opponent, Joe
Biden, drew from Democratic-leaning voters. Some referred to these group
of voters as Never Trump Republicans.

For this analysis, we study the counties in five battleground states
which together decided the election: Wisconsin, Michigan, Georgia,
Arizona, and Nevada. Biden won Georgia, Arizona, and Wisconsin by less
than a percentage point, and won Nevada and Michigan by less than 3
percentage points.

\begin{Shaded}
\begin{Highlighting}[]
\NormalTok{ds\_states }\OtherTok{\textless{}{-}}\NormalTok{ ds }\SpecialCharTok{|\textgreater{}} 
  \FunctionTok{filter}\NormalTok{(state }\SpecialCharTok{\%in\%} \FunctionTok{c}\NormalTok{(}\StringTok{"WISCONSIN"}\NormalTok{, }\StringTok{"MICHIGAN"}\NormalTok{, }\StringTok{"GEORGIA"}\NormalTok{, }\StringTok{"ARIZONA"}\NormalTok{, }\StringTok{"NEVADA"}\NormalTok{))}
\end{Highlighting}
\end{Shaded}

Recall that aggregate election results report how many votes Biden and
Trump received, but unlike cast vote records, they do not reveal which
of those votes came from Republicans and Democratic voters. Only cast
vote records can classify voters into partisan types based on how they
voted in all offices except President.

We first need to narrow down our data so that we only use voter-contest
pairs in contests contested by a Democrat and a Republican. In other
words, the voter needed to have a choice to vote for a Republican or
Democrat.

\begin{Shaded}
\begin{Highlighting}[]
\NormalTok{ds\_contested }\OtherTok{\textless{}{-}}\NormalTok{ ds\_states }\SpecialCharTok{|\textgreater{}} 
  \FunctionTok{collect}\NormalTok{() }\SpecialCharTok{|\textgreater{}} 
  \CommentTok{\# Contested contests}
  \FunctionTok{filter}\NormalTok{(}\FunctionTok{any}\NormalTok{(party }\SpecialCharTok{==} \StringTok{"REP"}\NormalTok{) }\SpecialCharTok{\&} \FunctionTok{any}\NormalTok{(party }\SpecialCharTok{==} \StringTok{"DEM"}\NormalTok{), }
         \AttributeTok{.by =} \FunctionTok{c}\NormalTok{(state, office, district)) }\SpecialCharTok{|\textgreater{}} 
  \CommentTok{\# Ballots with Presidential vote}
  \FunctionTok{filter}\NormalTok{(}\FunctionTok{any}\NormalTok{(office }\SpecialCharTok{==} \StringTok{"US PRESIDENT"}\NormalTok{), }
         \AttributeTok{.by =} \FunctionTok{c}\NormalTok{(state, county\_name, cvr\_id))}
\end{Highlighting}
\end{Shaded}

The first \texttt{filter()} command in this code limits to vote choices
for contested offices. For each state-office-district combination, we
examine if there are any Republican candidates \emph{and} any Democrats.
Contests that do not meet this criteria are dropped. 
We see here that users should use the combination of the variables \texttt{state}, \texttt{office}, and possibly \texttt{district} to identify contests. 
To identify a particular candidate, users should further use the variables \texttt{party} or \texttt{candidate}. 

The second
\texttt{filter()} command limits to ballots with a Presidential choice.
This excludes fragmented ballots where the President and the rest of the
ballot is separated. 
As shown in this command, users should work with the \texttt{case\_id} variable to identify a particular set of ballots. As discussed in the dataset description section, this ID is a numeric variable that is defined within counties, which are in turn unique within states (Two different counties in different states can have the same \texttt{county\_name} value).  
These numbers do not in any way indicate the time in which the
ballot was cast, or the personal identity of the voter. 
For more examples on how to extract such summaries from our data, see
Supplementary Information E.

Both commands are done after \texttt{collect()}
because the \texttt{arrow} package does not support group-specific
filter commands as of version 16.1.0.

We now construct a dataset where each row is a single voter. We first
create a dataset of Presidential votes:

\begin{Shaded}
\begin{Highlighting}[]
\DocumentationTok{\#\# Voters based on President}
\NormalTok{ds\_pres }\OtherTok{\textless{}{-}}\NormalTok{ ds\_contested }\SpecialCharTok{|\textgreater{}} 
  \FunctionTok{filter}\NormalTok{(office }\SpecialCharTok{==} \StringTok{"US PRESIDENT"}\NormalTok{) }\SpecialCharTok{|\textgreater{}} 
  \FunctionTok{select}\NormalTok{(}
\NormalTok{    state, county\_name, }
\NormalTok{    cvr\_id, candidate,}
    \AttributeTok{pres\_party =}\NormalTok{ party) }\SpecialCharTok{|\textgreater{}} 
  \FunctionTok{mutate}\NormalTok{(}\AttributeTok{pres =} \FunctionTok{case\_when}\NormalTok{(}
\NormalTok{    pres\_party }\SpecialCharTok{==} \StringTok{"REP"} \SpecialCharTok{\textasciitilde{}} \StringTok{"Trump"}\NormalTok{, }
\NormalTok{    pres\_party }\SpecialCharTok{==} \StringTok{"DEM"} \SpecialCharTok{\textasciitilde{}} \StringTok{"Biden"}\NormalTok{, }
\NormalTok{    pres\_party }\SpecialCharTok{==} \StringTok{"LBT"} \SpecialCharTok{\textasciitilde{}} \StringTok{"Libertarian"}\NormalTok{, }
\NormalTok{    candidate }\SpecialCharTok{==} \StringTok{"UNDERVOTE"} \SpecialCharTok{\textasciitilde{}} \StringTok{"Undervote"}\NormalTok{,}
    \AttributeTok{.default =} \StringTok{"Other"}\NormalTok{))}
\end{Highlighting}
\end{Shaded}

Separately, we construct a dataset that classifies the same voters based
on their non-Presidential vote choice. The variable
\texttt{nonpres\_party} is \texttt{Down-ballot\ Democrat} if the voter
only votes for Democrats down-ballot (using the \texttt{all()} command)
and it is \texttt{Down-ballot\ Republican} if the voter only votes for
Republicans down-ballot.

\begin{Shaded}
\begin{Highlighting}[]
\DocumentationTok{\#\# subset to all{-}Dem voters based on everything except President}
\NormalTok{ds\_D }\OtherTok{\textless{}{-}}\NormalTok{ ds\_contested }\SpecialCharTok{|\textgreater{}} 
  \FunctionTok{filter}\NormalTok{(office }\SpecialCharTok{!=} \StringTok{"US PRESIDENT"}\NormalTok{) }\SpecialCharTok{|\textgreater{}} 
  \FunctionTok{filter}\NormalTok{(}\FunctionTok{all}\NormalTok{(party }\SpecialCharTok{==} \StringTok{"DEM"}\NormalTok{), }\AttributeTok{.by =} \FunctionTok{c}\NormalTok{(state, county\_name, cvr\_id)) }\SpecialCharTok{|\textgreater{}} 
  \FunctionTok{distinct}\NormalTok{(state, county\_name, cvr\_id) }\SpecialCharTok{|\textgreater{}} 
  \FunctionTok{mutate}\NormalTok{(}\AttributeTok{nonpres\_party =} \StringTok{"Down{-}ballot Democrat"}\NormalTok{)}

\DocumentationTok{\#\# same subset, but for all{-}Rep voters}
\NormalTok{ds\_R }\OtherTok{\textless{}{-}}\NormalTok{ ds\_contested }\SpecialCharTok{|\textgreater{}} 
  \FunctionTok{filter}\NormalTok{(office }\SpecialCharTok{!=} \StringTok{"US PRESIDENT"}\NormalTok{) }\SpecialCharTok{|\textgreater{}} 
  \FunctionTok{filter}\NormalTok{(}\FunctionTok{all}\NormalTok{(party }\SpecialCharTok{==} \StringTok{"REP"}\NormalTok{), }\AttributeTok{.by =} \FunctionTok{c}\NormalTok{(state, county\_name, cvr\_id)) }\SpecialCharTok{|\textgreater{}} 
  \FunctionTok{distinct}\NormalTok{(state, county\_name, cvr\_id) }\SpecialCharTok{|\textgreater{}} 
  \FunctionTok{mutate}\NormalTok{(}\AttributeTok{nonpres\_party =} \StringTok{"Down{-}ballot Republican"}\NormalTok{)}
\end{Highlighting}
\end{Shaded}

Now we join voter's choices for President with their down-ballot
choices. Because each row is now a single voter, we join one-to-one
using \texttt{state}, \texttt{county\_name}, and \texttt{cvr\_id}.
Voters who were not classified into Democrats or Republican, are, by
construction, those who voted for some Democratic down-ballot candidates
and Republican down-ballot candidates, or undervoted in some of these
offices. We label them \texttt{Mixed} voters.

\begin{Shaded}
\begin{Highlighting}[]
\NormalTok{ds\_analysis }\OtherTok{\textless{}{-}}\NormalTok{ ds\_pres }\SpecialCharTok{|\textgreater{}}
  \FunctionTok{left\_join}\NormalTok{(}
    \FunctionTok{bind\_rows}\NormalTok{(ds\_D, ds\_R),}
    \AttributeTok{by =} \FunctionTok{c}\NormalTok{(}\StringTok{"state"}\NormalTok{, }\StringTok{"county\_name"}\NormalTok{, }\StringTok{"cvr\_id"}\NormalTok{), }\AttributeTok{relationship =} \StringTok{"one{-}to{-}one"}\NormalTok{) }\SpecialCharTok{|\textgreater{}}
  \FunctionTok{mutate}\NormalTok{(}\AttributeTok{nonpres\_party =} \FunctionTok{replace\_na}\NormalTok{(nonpres\_party, }\StringTok{"Mixed"}\NormalTok{))}
\end{Highlighting}
\end{Shaded}

\noindent Finally, we construct a cross-tabulation of this dataset using
the base-R \texttt{xtabs()} function.

\begin{Shaded}
\begin{Highlighting}[]
\FunctionTok{xtabs}\NormalTok{(}\SpecialCharTok{\textasciitilde{}}\NormalTok{ nonpres\_party }\SpecialCharTok{+}\NormalTok{ pres, ds\_analysis) }\SpecialCharTok{|\textgreater{}}
  \FunctionTok{addmargins}\NormalTok{()}
\end{Highlighting}
\end{Shaded}

\begin{verbatim}
                        pres
nonpres_party              Biden Libertarian   Other   Trump Undervote     Sum
  Down-ballot Democrat   3430497       15087   12012   43239      4478 3505313
  Down-ballot Republican   70136       26723   12396 3540120      9166 3658541
  Mixed                   793905       71431   21379  619255     16688 1522658
  Sum                    4294538      113241   45787 4202614     30332 8686512
\end{verbatim}

This table shows for example that among 3,505,313 solidly Democratic
voters, 3,430,497 voted for Joe Biden. We can show cell counts in terms
of proportions of the entire row, with the following base-R operation:

\begin{Shaded}
\begin{Highlighting}[]
\NormalTok{xtprop }\OtherTok{\textless{}{-}} \FunctionTok{xtabs}\NormalTok{(}\SpecialCharTok{\textasciitilde{}}\NormalTok{ nonpres\_party }\SpecialCharTok{+}\NormalTok{ pres, ds\_analysis) }\SpecialCharTok{|\textgreater{}}
  \FunctionTok{prop.table}\NormalTok{(}\AttributeTok{margin =} \DecValTok{1}\NormalTok{) }\SpecialCharTok{|\textgreater{}}
  \FunctionTok{round}\NormalTok{(}\DecValTok{3}\NormalTok{)}

\DocumentationTok{\#\# add margins}
\NormalTok{N }\OtherTok{\textless{}{-}} \FunctionTok{xtabs}\NormalTok{(}\SpecialCharTok{\textasciitilde{}}\NormalTok{ nonpres\_party, ds\_analysis) }\SpecialCharTok{|\textgreater{}}
  \FunctionTok{format}\NormalTok{(}\AttributeTok{big.mark =} \StringTok{","}\NormalTok{)}

\DocumentationTok{\#\# reorder columns and append totals}
\NormalTok{xtprop[, }\FunctionTok{c}\NormalTok{(}\StringTok{"Biden"}\NormalTok{, }\StringTok{"Trump"}\NormalTok{, }\StringTok{"Libertarian"}\NormalTok{, }\StringTok{"Undervote"}\NormalTok{)] }\SpecialCharTok{|\textgreater{}}
  \FunctionTok{cbind}\NormalTok{(N) }\SpecialCharTok{|\textgreater{}}
\NormalTok{  kableExtra}\SpecialCharTok{::}\FunctionTok{kbl}\NormalTok{(}\AttributeTok{format =} \StringTok{"latex"}\NormalTok{, }\AttributeTok{booktabs =} \ConstantTok{TRUE}\NormalTok{)}
\end{Highlighting}
\end{Shaded}

\begin{table}[!h]
\begin{tabular}[t]{llllll}
\toprule
  & Biden & Trump & Libertarian & Undervote & N\\
\midrule
Down-ballot Democrat & 0.979 & 0.012 & 0.004 & 0.001 & 3,505,313\\
Down-ballot Republican & 0.019 & 0.968 & 0.007 & 0.003 & 3,658,541\\
Mixed & 0.521 & 0.407 & 0.047 & 0.011 & 1,522,658\\
\bottomrule
\end{tabular}
\caption{\textbf{Party Loyalty in Five Battle Ground States}}
\label{tab:party-loyalty}
\end{table}

\bigskip

This formatted table (\autoref{tab:party-loyalty}) shows more clearly that the ticket splitting rate
among solid partisans was on the order of 1 percent in this sample. Such
small subgroups are almost impossible to detect in a survey. In
contrast, 97 percent of solid Republicans stuck with their party's
nominee, Trump, and 98 percent of solid Democrats stuck with Biden.
Trump's party loyalty was a percentage point smaller than Biden's.

A starker difference arises in the mixed group (those who vote for some
Republicans and some Democrats, or undervoted, down-ballot). Biden won
this group of weak partisans by more than 10 points. Close to 5 percent
of this group voted for the third-party Libertarian candidate for
President, instead of picking either Biden or Trump. Undervoting for
President was low, less than 1 percent, in these group of voters.

More can be done to examine if these results vary by state, county, or
precinct. Future versions of this dataset can also include ballot
measures and local candidates that give more context of these patterns. 
\newpage
\FloatBarrier

\section*{Code availability}

The code we use to construct the dataset is available at \url{https://github.com/kuriwaki/cvr_harvard-mit_scripts}. To parse some JSON cast vote records, we use the \texttt{dominionCVR} R library by Kuriwaki and Lewis (\url{https://github.com/kuriwaki/dominionCVR}).

\bibliography{refs.bib}

\begin{thebibliography}{10}
\urlstyle{rm}
\expandafter\ifx\csname url\endcsname\relax
  \def\url#1{\texttt{#1}}\fi
\expandafter\ifx\csname urlprefix\endcsname\relax\def\urlprefix{URL }\fi
\expandafter\ifx\csname doiprefix\endcsname\relax\def\doiprefix{DOI: }\fi
\providecommand{\bibinfo}[2]{#2}
\providecommand{\eprint}[2][]{\url{#2}}

\bibitem{road97}
\bibinfo{author}{King, G.} \emph{et~al.}
\newblock \bibinfo{title}{{The Record of American Democracy, 1984-1990}}
  (\bibinfo{year}{1997}).
\newblock
  \bibinfo{note}{\url{https://road.hmdc.harvard.edu/pages/road-documentation}}.

\bibitem{ansolabehere14}
\bibinfo{author}{Ansolabehere, S.}, \bibinfo{author}{Palmer, M.} \&
  \bibinfo{author}{Lee, A.}
\newblock \bibinfo{title}{{Precinct-Level Election Data, 2002-2012}},
  \url{https://doi.org/10.7910/DVN/YN4TLR} (\bibinfo{year}{2014}).

\bibitem{precincts22}
\bibinfo{author}{Baltz, S.} \emph{et~al.}
\newblock \bibinfo{journal}{\bibinfo{title}{American election results at the
  precinct level}}.
\newblock {\emph{\JournalTitle{Nature Scientific Data}}}
  \url{https://doi.org/10.1038/s41597-022-01745-0} (\bibinfo{year}{2022}).

\bibitem{VEST:2022}
\bibinfo{author}{{Voting and Election Science Team}}.
\newblock \bibinfo{title}{{Precinct-Level Election Results}},
  \url{https://doi.org/10.7910/DVN/K7760H} (\bibinfo{year}{2024}).
\newblock \bibinfo{note}{Last accessed 2024}.

\bibitem{wack2019cast}
\bibinfo{author}{Wack, J.~P.}
\newblock \bibinfo{journal}{\bibinfo{title}{Cast vote records common data
  format specification version 1.0}}.
\newblock {\emph{\JournalTitle{National Institute of Standards and
  Technology}}} \url{https://doi.org/10.6028/NIST.SP.1500-103}
  (\bibinfo{year}{2019}).

\bibitem{dataverse}
\bibinfo{author}{Kuriwaki, S.}, \bibinfo{author}{Reece, M.} \emph{et~al.}
\newblock \bibinfo{title}{{Cast vote records: A database of ballots from the
  2020 U.S. Election, Harvard Dataverse}},
  \url{https://doi.org/10.7910/DVN/PQQ3KV} (\bibinfo{year}{2024}).

\bibitem{green2023foia}
\bibinfo{author}{Green, R.}
\newblock \bibinfo{journal}{\bibinfo{title}{{FOIA-Flooded Elections}}}.
\newblock {\emph{\JournalTitle{Ohio State Law Journal}}}
  \bibinfo{pages}{255--306} (\bibinfo{year}{2024}).

\bibitem{leingang2022}
\bibinfo{author}{Leingang, R.}
\newblock \bibinfo{journal}{\bibinfo{title}{Election activists are seeking the
  ``cast vote record'' from 2020. here's what it is and why they want it.}}
\newblock {\emph{\JournalTitle{Votebeat, September 7, 2022}}}
  (\bibinfo{year}{2022}).
\newblock \bibinfo{note}{\url{https://perma.cc/55F4-NVKW}}.

\bibitem{odonnell}
\bibinfo{author}{O'Donnell, J.}
\newblock \bibinfo{journal}{\bibinfo{title}{The fingerprints of fraud: Evidence
  and analysis of multi-state conspiracy to defraud the 2020 general election,
  vol. 1}}.
\newblock {\emph{\JournalTitle{Unpublished Manuscript Available Online}}}
  (\bibinfo{year}{2023}).
\newblock \bibinfo{note}{\url{https://perma.cc/6KY3-3GFK}}.

\bibitem{bloombergnews2022}
\bibinfo{author}{{Bloomberg Technology}}.
\newblock \bibinfo{title}{{`Raccoon Army' Swamps Election Officials in Dubious
  Campaign to Disprove Results}} (\bibinfo{year}{2022}).
\newblock \bibinfo{note}{October 25, 2022, \url{https://perma.cc/J2H6-TAUD}}.

\bibitem{grimmer2024evaluating}
\bibinfo{author}{Grimmer, J.}, \bibinfo{author}{Herron, M.~C.} \&
  \bibinfo{author}{Tyler, M.}
\newblock \bibinfo{journal}{\bibinfo{title}{Evaluating a new generation of
  expansive claims about vote manipulation}}.
\newblock {\emph{\JournalTitle{Election Law Journal}}}
  \url{https://doi.org/elj.2022.0070} (\bibinfo{year}{2024}).

\bibitem{burden2009split}
\bibinfo{author}{Burden, B.~C.} \& \bibinfo{author}{Kimball, D.~C.}
\newblock \emph{\bibinfo{title}{{Why Americans Split their Tickets: Campaigns,
  Competition, and Divided Government}}} (\bibinfo{publisher}{University of
  Michigan Press}, \bibinfo{year}{2009}).

\bibitem{rogers2016national}
\bibinfo{author}{Rogers, S.}
\newblock \bibinfo{journal}{\bibinfo{title}{National forces in state
  legislative elections}}.
\newblock {\emph{\JournalTitle{The Annals of the American Academy of Political
  and Social Science}}} \textbf{\bibinfo{volume}{667}},
  \bibinfo{pages}{207--225} (\bibinfo{year}{2016}).

\bibitem{malzhanhall}
\bibinfo{author}{Malzhan, J.} \& \bibinfo{author}{Hall, A.~B.}
\newblock \bibinfo{journal}{\bibinfo{title}{Election-denying republican
  candidates underperformed in the 2022 midterms}}.
\newblock {\emph{\JournalTitle{Forthcoming, American Political Science
  Review}}} \url{https://doi.org/10.1017/S0003055424001084}
  (\bibinfo{year}{2024}).

\bibitem{reece2024hidden}
\bibinfo{author}{Reece, M.} \emph{et~al.}
\newblock \bibinfo{journal}{\bibinfo{title}{{Hidden Partisanship in American
  Elections}}}.
\newblock {\emph{\JournalTitle{Preprint}}}  (\bibinfo{year}{2024}).
\newblock \bibinfo{note}{\url{https://ssrn.com/abstract=4721012}}.

\bibitem{ConevskaMPSA2024}
\bibinfo{author}{Conevska, A.} \emph{et~al.}
\newblock \bibinfo{journal}{\bibinfo{title}{{How Partisan are U.S. Local
  Elections? Evidence from 2020 Cast Vote Records}}}.
\newblock {\emph{\JournalTitle{Preprint}}}  (\bibinfo{year}{2024}).

\bibitem{kuriwaki_JMP}
\bibinfo{author}{Kuriwaki, S.}
\newblock \bibinfo{journal}{\bibinfo{title}{{Ticket Splitting in a Nationalized
  Era}}}.
\newblock {\emph{\JournalTitle{Preprint}}}
  \url{https://doi.org/10.31235/osf.io/bvgz3} (\bibinfo{year}{2023}).

\bibitem{dubin1992patterns}
\bibinfo{author}{Dubin, J.~A.} \& \bibinfo{author}{Gerber, E.~R.}
\newblock \bibinfo{title}{Patterns of voting on ballot propositions: A mixture
  model of voter types}.
\newblock \bibinfo{type}{Tech. Rep.}, \bibinfo{institution}{California
  Institute of Technology, Division of the Humanities and Social Sciences}
  (\bibinfo{year}{1992}).
\newblock \url{https://doi.org/10.7907/xprd7-7jt08}.

\bibitem{Gerber2004}
\bibinfo{author}{Gerber, E.~R.} \& \bibinfo{author}{Lewis, J.~B.}
\newblock \bibinfo{journal}{\bibinfo{title}{{Beyond the Median: Voter
  Preferences, District Heterogeneity, and Political Representation}}}.
\newblock {\emph{\JournalTitle{Journal of Political Economy}}}
  \textbf{\bibinfo{volume}{112}}, \bibinfo{pages}{1364--1383},
  \url{https://doi.org/10.1086/424737} (\bibinfo{year}{2004}).

\bibitem{besley2008issue}
\bibinfo{author}{Besley, T.} \& \bibinfo{author}{Coate, S.}
\newblock \bibinfo{journal}{\bibinfo{title}{{Issue Unbundling via Citizens'
  Initiatives}}}.
\newblock {\emph{\JournalTitle{Quarterly Journal of Political Science}}}
  \textbf{\bibinfo{volume}{3}}, \bibinfo{pages}{379--397}
  (\bibinfo{year}{2008}).

\bibitem{stromberg2008electoral}
\bibinfo{author}{Str{\"o}mberg, D.}
\newblock \bibinfo{journal}{\bibinfo{title}{{How the Electoral College
  influences campaigns and policy: the probability of being Florida}}}.
\newblock {\emph{\JournalTitle{American Economic Review}}}
  \textbf{\bibinfo{volume}{98}}, \bibinfo{pages}{769--807}
  (\bibinfo{year}{2008}).

\bibitem{larcinese2013testing}
\bibinfo{author}{Larcinese, V.}, \bibinfo{author}{Snyder, J.~M.} \&
  \bibinfo{author}{Testa, C.}
\newblock \bibinfo{journal}{\bibinfo{title}{Testing models of distributive
  politics using exit polls to measure voters’ preferences and
  partisanship}}.
\newblock {\emph{\JournalTitle{British Journal of Political Science}}}
  \textbf{\bibinfo{volume}{43}}, \bibinfo{pages}{845--875}
  (\bibinfo{year}{2013}).

\bibitem{hill2017changing}
\bibinfo{author}{Hill, S.~J.}
\newblock \bibinfo{journal}{\bibinfo{title}{{Changing votes or changing voters?
  How candidates and election context swing voters and mobilize the base}}}.
\newblock {\emph{\JournalTitle{Electoral Studies}}}
  \textbf{\bibinfo{volume}{48}}, \bibinfo{pages}{131--148}
  (\bibinfo{year}{2017}).

\bibitem{dowling2024SC}
\bibinfo{author}{Dowling, C.~M.}, \bibinfo{author}{Miller, M.~G.} \&
  \bibinfo{author}{Morris, K.}
\newblock \bibinfo{journal}{\bibinfo{title}{{Can Voters Locate Copartisan
  Candidates in Nonpartisan Elections? Evidence from Cast Vote Records}}}.
\newblock {\emph{\JournalTitle{Working Paper}}}  (\bibinfo{year}{2024}).

\bibitem{DeLuca2022}
\bibinfo{author}{DeLuca, K.}
\newblock \bibinfo{journal}{\bibinfo{title}{Editor's choice: Measuring
  candidate quality using local newspaper endorsements}}.
\newblock {\emph{\JournalTitle{APSA Preprints}}}
  \url{https://doi.org/10.33774/apsa-2023-3qdmj-v3} (\bibinfo{year}{2024}).

\bibitem{alvarez2018low}
\bibinfo{author}{Alvarez, R.~M.}, \bibinfo{author}{Hall, T.~E.} \&
  \bibinfo{author}{Levin, I.}
\newblock \bibinfo{journal}{\bibinfo{title}{{Low-information voting: Evidence
  from instant-runoff elections}}}.
\newblock {\emph{\JournalTitle{American Politics Research}}}
  \textbf{\bibinfo{volume}{46}}, \bibinfo{pages}{1012--1038}
  (\bibinfo{year}{2018}).

\bibitem{atsusaka2024analyzing}
\bibinfo{author}{Atsusaka, Y.}
\newblock \bibinfo{journal}{\bibinfo{title}{Analyzing ballot order effects when
  voters rank candidates}}.
\newblock {\emph{\JournalTitle{Political Analysis}}} \bibinfo{pages}{1--9}
  (\bibinfo{year}{2024}).

\bibitem{Herron2007}
\bibinfo{author}{Herron, M.~C.} \& \bibinfo{author}{Lewis, J.~B.}
\newblock \bibinfo{journal}{\bibinfo{title}{{Did Ralph Nader Spoil a Gore
  Presidency? A Ballot-level Study of Green and Reform Party Voters in the 2000
  Presidential Election}}}.
\newblock {\emph{\JournalTitle{Quarterly Journal of Political Science}}}
  \textbf{\bibinfo{volume}{2}}, \bibinfo{pages}{205--226},
  \url{https://doi.org/10.1561/100.00005039} (\bibinfo{year}{2007}).

\bibitem{adler13}
\bibinfo{author}{Adler, E.~S.} \& \bibinfo{author}{Hall, T.~E.}
\newblock \bibinfo{journal}{\bibinfo{title}{Ballots, transparency, and
  democracy}}.
\newblock {\emph{\JournalTitle{Election Law Journal}}}
  \textbf{\bibinfo{volume}{12}}, \bibinfo{pages}{146--161},
  \url{https://doi.org/10.1089/elj.2012.0179} (\bibinfo{year}{2013}).

\bibitem{bernhard2017public}
\bibinfo{author}{Bernhard, M.} \emph{et~al.}
\newblock \bibinfo{title}{Public evidence from secret ballots}.
\newblock In \emph{\bibinfo{booktitle}{Electronic Voting: Second International
  Joint Conference, E-Vote-ID 2017, Bregenz, Austria, October 24-27, 2017,
  Proceedings 2}}, \bibinfo{pages}{84--109},
  \url{https://doi.org/10.1007/978-3-319-68687-5_6}
  (\bibinfo{organization}{Springer}, \bibinfo{year}{2017}).

\bibitem{gerber2013there}
\bibinfo{author}{Gerber, A.~S.}, \bibinfo{author}{Huber, G.~A.},
  \bibinfo{author}{Doherty, D.} \& \bibinfo{author}{Dowling, C.~M.}
\newblock \bibinfo{journal}{\bibinfo{title}{Is there a secret ballot? ballot
  secrecy perceptions and their implications for voting behaviour}}.
\newblock {\emph{\JournalTitle{British Journal of Political Science}}}
  \textbf{\bibinfo{volume}{43}}, \bibinfo{pages}{77--102},
  \url{https://doi.org/10.1017/S000712341200021X} (\bibinfo{year}{2013}).

\bibitem{atkeson23}
\bibinfo{author}{Atkeson, L.~R.}, \bibinfo{author}{McKown-Dawson, E.},
  \bibinfo{author}{{Hood III}, M.} \& \bibinfo{author}{Stein, R.}
\newblock \bibinfo{journal}{\bibinfo{title}{Voter perceptions of secrecy in the
  2020 election}}.
\newblock {\emph{\JournalTitle{Election Law Journal}}}
  \textbf{\bibinfo{volume}{22}}, \bibinfo{pages}{234--253},
  \url{https://doi.org/10.1089/elj.2022.0064} (\bibinfo{year}{2023}).

\bibitem{jaffe2023effect}
\bibinfo{author}{Jaffe, J.}, \bibinfo{author}{Loffredo, J.},
  \bibinfo{author}{Baltz, S.}, \bibinfo{author}{Flores, A.} \&
  \bibinfo{author}{Stewart~III, C.}
\newblock \bibinfo{journal}{\bibinfo{title}{Trust in the count: Improving voter
  confidence with post-election audits}}.
\newblock {\emph{\JournalTitle{Public Opinion Quarterly}}}
  \textbf{\bibinfo{volume}{88}}, \bibinfo{pages}{585–607}
  (\bibinfo{year}{2024}).

\bibitem{wand2001butterfly}
\bibinfo{author}{Wand, J.~N.} \emph{et~al.}
\newblock \bibinfo{journal}{\bibinfo{title}{{The butterfly did it: The aberrant
  vote for Buchanan in Palm Beach County, Florida}}}.
\newblock {\emph{\JournalTitle{American Political Science Review}}}
  \textbf{\bibinfo{volume}{95}}, \bibinfo{pages}{793--810}
  (\bibinfo{year}{2001}).

\bibitem{Bafumi2012}
\bibinfo{author}{Bafumi, J.}, \bibinfo{author}{Herron, M.~C.},
  \bibinfo{author}{Hill, S.~J.} \& \bibinfo{author}{Lewis, J.~B.}
\newblock \bibinfo{journal}{\bibinfo{title}{{Alvin Greene? Who? How Did He Win
  the United States Senate Nomination in South Carolina?}}}
\newblock {\emph{\JournalTitle{Election Law Journal}}}
  \textbf{\bibinfo{volume}{11}}, \bibinfo{pages}{358--379},
  \url{https://doi.org/10.1089/elj.2011.0137} (\bibinfo{year}{2012}).

\bibitem{biggers2023can}
\bibinfo{author}{Biggers, D.~R.} \emph{et~al.}
\newblock \bibinfo{journal}{\bibinfo{title}{Can addressing integrity concerns
  about mail balloting increase turnout? results from a large-scale field
  experiment in the 2020 presidential election}}.
\newblock {\emph{\JournalTitle{Journal of Experimental Political Science}}}
  \textbf{\bibinfo{volume}{10}}, \bibinfo{pages}{413--425},
  \url{https://doi.org/10.1017/XPS.2022.31} (\bibinfo{year}{2023}).

\bibitem{gerber2013perceptions}
\bibinfo{author}{Gerber, A.~S.}, \bibinfo{author}{Huber, G.~A.},
  \bibinfo{author}{Doherty, D.}, \bibinfo{author}{Dowling, C.~M.} \&
  \bibinfo{author}{Hill, S.~J.}
\newblock \bibinfo{journal}{\bibinfo{title}{Do perceptions of ballot secrecy
  influence turnout? results from a field experiment}}.
\newblock {\emph{\JournalTitle{American Journal of Political Science}}}
  \textbf{\bibinfo{volume}{57}}, \bibinfo{pages}{537--551},
  \url{https://doi.org/10.1111/ajps.12019} (\bibinfo{year}{2013}).

\bibitem{williams2024votes}
\bibinfo{author}{Williams, J.~R.}, \bibinfo{author}{Baltz, S.} \&
  \bibinfo{author}{Stewart, C.}
\newblock \bibinfo{journal}{\bibinfo{title}{Votes can be confidently bought in
  some ranked ballot elections, and what to do about it}}.
\newblock {\emph{\JournalTitle{Political Analysis}}} \bibinfo{pages}{1--13},
  \url{https://doi.org/10.1017/pan.2024.4} (\bibinfo{year}{2024}).

\bibitem{kuriwaki23}
\bibinfo{author}{Kuriwaki, S.}, \bibinfo{author}{Lewis, J.~B.} \&
  \bibinfo{author}{Morse, M.}
\newblock \bibinfo{journal}{\bibinfo{title}{The still secret ballot: The
  limited privacy cost of transparent election results}}.
\newblock {\emph{\JournalTitle{Preprint}}}
  \url{https://doi.org/10.48550/arXiv.2308.04100} (\bibinfo{year}{2023}).

\bibitem{medsl2020bystate}
\bibinfo{author}{{MIT Election Data and Science Lab}}.
\newblock \bibinfo{title}{Precinct-level returns 2020 by individual state}.
\newblock \bibinfo{howpublished}{Harvard Dataverse},
  \url{https://doi.org/10.7910/DVN/NT66Z3} (\bibinfo{year}{2022}).

\bibitem{FEC2020}
\bibinfo{author}{{Federal Election Commission}}.
\newblock \bibinfo{title}{{Federal Elections 2020: Election Results for the
  U.S. President, the U.S. Senate and the U.S. House of Representatives}}.
\newblock \bibinfo{type}{Tech. Rep.}, \bibinfo{institution}{United States
  Government} (\bibinfo{year}{2022}).
\newblock \bibinfo{note}{\url{https://perma.cc/DE4X-SQFC}}.

\bibitem{mcdonnell2021}
\bibinfo{author}{{Dane County Clerk}}.
\newblock \bibinfo{title}{Re: November 2020 general election ballot images}.
\newblock \bibinfo{howpublished}{Letter to Brian McGrath and Kyle Koenen on
  August 13, 2021} (\bibinfo{year}{2021}).

\bibitem{bpcturnover}
\bibinfo{author}{Ferrer, J.}, \bibinfo{author}{Thompson, D.~M.} \&
  \bibinfo{author}{Orey, R.}
\newblock \bibinfo{journal}{\bibinfo{title}{Election official turnover rates
  from 2000-2024}}.
\newblock {\emph{\JournalTitle{Bipartisan Policy Center}}}
  (\bibinfo{year}{2024}).
\newblock
  \bibinfo{note}{\url{https://bipartisanpolicy.org/report/election-official-turnover-rates-from-2000-2024/}}.

\end{thebibliography}

\section*{Acknowledgements}

We thank Luka Bulić Bračulj for additional work analyzing the data. We also thank Scott McDonnell of Dane County for answering our questions about Dane County, as well as other local election officials that answered our questions.
We also thank Larry Moore for inspiration on the usage example.
This material is based upon work supported by the National Science Foundation Graduate Research Fellowship Program under Grant No. DGE-2139841 (to T.S.).

\section*{Author contributions statement}

S.K. drafted the manuscript; 
M.R., J.S., and J.B.L. were the principal maintainers of the database construction;
A.C., C.S., J.B.L., J.R.L., J.S., M.R., S.H., S.K. contributed to the manuscript;
A.C., C.M., J.B.L., J.R.L., J.S., K.A., K.M., S.H., S.K., M.R., T.S., Z.G., performed data processing, validation, investigation;
J.B.L., J.R.L., J.S., M.R., S.K., S.B., T.S., wrote software;
C.S., J.S., M.R., S.B., S.K. supervised the project.
All authors reviewed the paper.

\section*{Competing interests} 

Authors declare no conflicting interests that might be perceived to influence the results and/or discussion reported in this paper.
This project was determined as not human subjects research by the MIT and Yale Institutional Review Board (IRB).

\clearpage

\begin{center}
{\LARGE \textsf{\textbf{Supplementary Information for}}}

\medskip

{\large \textsf{\textbf{``Cast vote records: A database of ballots from the 2020 U.S. Election''}}}
\end{center}

\appendix
\renewcommand{\thetable}{S\arabic{table}}
\renewcommand{\thefigure}{S\arabic{figure}}
\renewcommand{\thesection}{\Alph{section}.}

\setcounter{table}{0}
\setcounter{figure}{0}

\noindent This document is supplementary information for the Cast Vote Records Data Descriptor\cite{dataverse}.

\section{Fixing Fragmented Ballots} \label{sec:fusing-pages}

The counties in \autoref{tab:pagination-counties} did not have a clear cast vote record identifier. We nevertheless paired records as one voter (one CVR ID) using a matching algorithm described in Algorithm \ref{alg:pagination}. The algorithm relies on the user to supply two arguments, (1) a grouping column, \texttt{args.groupcol}, that is a condition for pages to belong to the same voter and (2) a target column, \texttt{args.targetcol}, in the CVR that can be used to match pages together. For example, if the column is the President, if two rows actually belong to the same voter,  the pair must have an identical value for the \texttt{args.groupcol}, and one row will have a value for President and the other row will be blank for President.

\FloatBarrier
\begin{table}
\caption{\textbf{Counties Where Algorithm \ref{alg:pagination} was used to re-connect pages}}
\label{tab:pagination-counties}
\centering
\begin{tabular}{ll}
\toprule
County & Group Column \\\midrule
Alameda, California & Ballot Type \\
Contra Costa, California & Ballot Type\\
Kings, California & Ballot Type \\
Merced, California & Ballot Style\\
Riverside, California & Ballot Type \\
San\ Benito, California & Ballot Type \\
San\ Mateo, California & Ballot Type\\
Sonoma, California &  Ballot Type\\
Yuba, California & Ballot Type\\
Denver, Colorado & Ballot Type\\
Eagle, Colorado &  Ballot Type\\
Routt, Colorado &  Ballot Type\\
Gwinnett, Georgia & Ballot Type\\
Baltimore, Maryland & Ballot Style\\
Baltimore City, Maryland & Ballot Style\\
Montgomery, Maryland & Ballot Style\\
Prince George's Maryland & Ballot Style\\
Butler, Ohio & Ballot Type\\
Champaign, Ohio & BallotStyleID\\
Cuyahoga, Ohio & Ballot Style \\
Greene, Ohio & Ballot Type\\
Rhode\ Island & Ballot Style\\\bottomrule
\end{tabular}
\end{table}

\begin{algorithm}
\caption{Process Missing Values in Dataset}
\label{alg:pagination}
\begin{algorithmic}[1]
\State Initialize an empty set \texttt{used\_rows} to keep track of processed rows.
\State Determine the total number of rows in the dataset, \texttt{num\_rows}.
\For{each row \texttt{current\_row} at index $i$ in the dataset}
    \If{$i$ is in \texttt{used\_rows}}
        \State \textbf{continue}
    \EndIf
    \If{\texttt{current\_row[args.targetcol]} is missing a value}
        \For{each index $j$ in expanding spiral around $i$}
            \If{$j \geq \texttt{num\_rows}$}
                \State \textbf{break}
            \EndIf
            \If{$j$ is not in \texttt{used\_rows} and \texttt{current\_row[args.groupcol]} equals \texttt{data[j, args.groupcol]} and \texttt{data[j, args.targetcol]} is not missing}
                \State Merge \texttt{current\_row} with the complementary row at index $j$ to form \texttt{merged\_row}.
                \State Update \texttt{data[i]} with \texttt{merged\_row}.
                \State Add $j$ to \texttt{used\_rows}.
                \State \textbf{break}
            \EndIf
        \EndFor
    \EndIf
\EndFor
\end{algorithmic}
\end{algorithm}

\FloatBarrier
\clearpage

\section{Precinct Name Standardization}
\label{sec:precinct-matching}

Linking CVR data to other precinct-level data sources is not straightforward. The released CVR data typically contains information about the voting precinct in which each ballot was cast. However, there is no agreed-upon standard for the labeling of precincts. The precinct labels used in the CVR data often do not match those used in the official precinct-level results published by the county or state. Further, CVRs sometimes provide more detailed geographic information (commonly called ``subprecinct'' information) that is not included in the official precinct-level tallies.

In order to make the CVR data more useful to researchers, we manually constructed a crosswalk from the precinct labels used in the CVR data to the precinct labels in a widely-used freely-available national database of certified precinct-level 2020 General Election results published by the MIT Election Data and Science Lab (see the descriptor by Baltz \textit{et al.}\cite{precincts22}). Using that crosswalk, we have appended the MEDSL precinct label to our dataset. For all linkages that we include in the data, we are completely confident that we have correctly matched precincts in the CVR and the MEDSL data. The code to construct the crosswalk can be found in our main codebase, currently at \url{https://github.com/kuriwaki/cvr_harvard-mit_scripts/tree/main/code/01_build-returns/code}.

We created this crosswalk by the following steps:
\begin{enumerate}
    \item First, we aggregated the CVR data to the (sub)precinct level based on the precinct labels that they include. (Note, for counties such as Los Angeles, California for which the CVRs do not provide precinct identifiers, we have not attempted to place the CVRs in MEDSL precincts. Such a linkage is not in general possible, though it might be accomplished in some cases by using the set of contests included on each ballot -- sometimes called the ``ballot style'' -- to identify the precinct.)  We then attempted to match our CVR-based precinct records to the MEDSL precinct data.
    Working county-by-county, we matched CVR precincts to MEDSL precincts based upon the reported number of votes cast for Democratic and Republican candidates for US President, US Senate, US House, State Senate, and State House. For many counties, unique exact matches for every CVR precinct could be found among the MEDSL precincts. For those counties, this set of unique exact matches was used as the crosswalk between CVR and MEDSL precincts labels.
    \item For counties in which unique exact candidate vote matches could not be found for every precinct, we transformed the CVR precinct names to more closely match the format of the MEDSL precinct names. We then disambiguated cases in which a given CVR precinct's candidate vote total exactly matched more than one MEDSL precinct using the edit distance between the transformed CVR precinct name and the MEDSL precinct name by selecting the potential match having the smallest edit distance. In cases in which no exact match based on the candidate vote totals existed, we established a linkage if the transformed CVR precinct name exactly matched the MEDSL precinct name.
    \item For those counties containing precincts for which neither the candidate vote totals nor the transformed precinct names could be exactly matched across the CVR and MEDSL data, potential matches were identified by the smallest sum of absolute deviations in candidate votes and confirmed by manual comparison of the CVR and the MEDSL precinct labels. Where this was not possible, no linkage was established.
    \item For some counties, we determined that the CVR precinct labels included subprecinct information meaning that the CVR-based precinct-level data was at a lower level of aggregation than that reported in the MEDSL data. In those counties, we transformed to CVR precinct names to remove the subprecinct information and re-aggregated to produce CVR precinct records at the same level of aggregation as the MEDSL precincts before constructing the crosswalk using the methods described above.
\end{enumerate}

MEDSL precinct labels were not applied to records for which a precinct identifier was not present in the CVR data nor to CVR precincts for which neither the names nor the vote totals established a convincing linkage between the two data sources.

In four Michigan Counties (Alcona, Clinton, Gladwin, and Missaukee) few or none of the CVR precincts could be matched to MEDSL precincts using the method described. In these four counties, the inability to link CVR precincts to MEDSL precincts appeared to be the result of irregularities in the MEDSL data. Those issues may be resolved in the future.

Fifteen other counties contained between one and five CVR precincts that could not be linked to MEDSL precincts. All of the precincts in the remaining 296 counties for which CVR precinct information was available could be linked to MEDSL precincts. In total, 28,540 CVR unique (sub)precincts were linked to 23,467 MEDSL precincts.

\section{Third Party Candidates and Write-ins}
\label{sec:greenparty}

Designations for third parties vary widely by state. This is compounded by the different ways counties print party on the ballot, and the types of voting machines store and record information about third party candidates and write-ins.

Howie Hawkins was the Green Party's nominee for U.S. President but the Green Party did not get ballot access in some states.
Among the states that we examined, the Green Party appears to not have had ballot access in Arizona, Georgia, Pennsylvania, and Wisconsin. Some jurisdictions, e.g., in Wisconsin, do not report out Hawkins as a specific candidate in either the cast vote record or the election returns; he is lumped into simply \texttt{WRITEIN}. In Georgia, the cast vote record and the election returns report Hawkins as its own candidate, but they are still write-ins.  (See \url{https://ballotpedia.org/Georgia\_official\_sample\_ballots,\_2020} for a sample ballot, and \url{https://bit.ly/3RYhL3A} for an example of how Hawkins is reported in an election return.)  
In all these cases, the \texttt{party} value for Hawkins is a Write-in, not the Green Party.
In contrast, New Jersey's CVRs records each and every write-in choice as a valid candidate.

It is worth noting that some of the data on \texttt{votedatabase.com} lost information about write-ins due to information loss from Excel to CSV.
In some machines, notably the ES\&S DS200 scanner, write-ins are scanned and can produce a cast vote record in Microsoft Excel. In these sheets, the write-in candidates are not transcribed into text but stored as an image file. However, O'Donnell instructed his collaborators to upload these data as a plain-text CSV file, which loses information in the image. In these cases, write-in votes are tracked as blank cells and excluded from our dataset. Our data, therefore, systematically misses write-in votes in many counties that use this format of cast vote record.

\section{Included Counties} \label{sec:counties}

\begin{figure}[tbp]
\caption{\textbf{Map of Counties Included}. \emph{(a) displays the included counties on a map of all counties in the United States. Counties shaded black are included in the data, counties shaded dark gray are part of a state that has other counties present in the data, but are not themselves present, and counties shaded light gray are neither present in the data nor is any other county from their state. (b) highlights only the states from which our data includes any counties, for easier viewing. Note that the visual area of each county is not representative of its population.}}
\label{fig:counties-map}
\includegraphics[width=\linewidth]{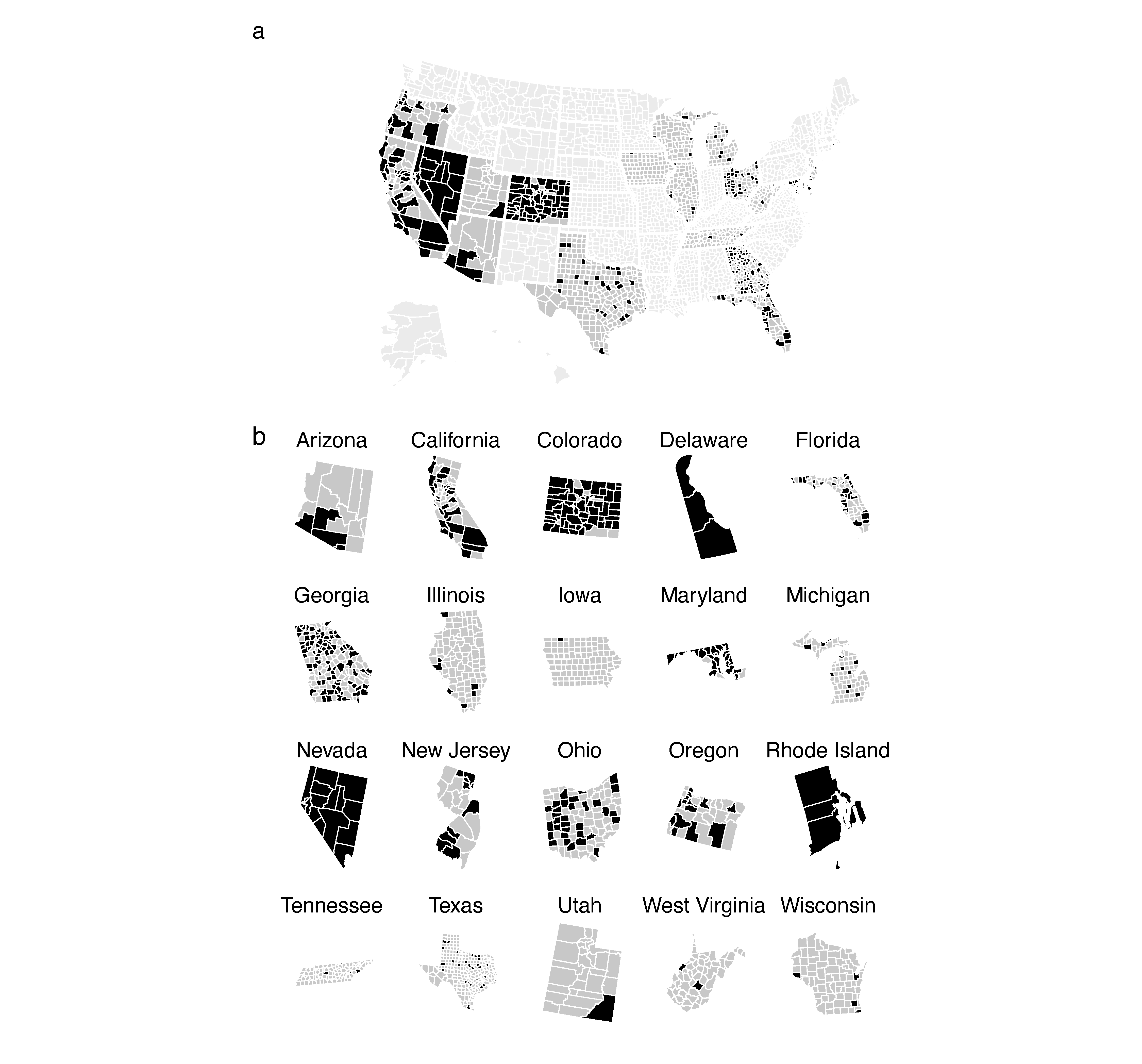}
\end{figure}

See \texttt{county\_info.xlsx} in the Dataverse repository for the list of counties included in our data release. 
\autoref{fig:counties-map} shows the geographic location of these counties. 

\begin{figure}[tbp]
\caption{\textbf{Coverage of CVR data}. \emph{The percentage shows the fraction of total voters contained in the CVR for that state and office. States are not shown on the map if the office was not on the ballot. States have no percentage if CVRs were not available for that state.}}
\label{fig:map-frac}
\begin{adjustwidth}{0cm}{0cm}
\includegraphics[width=\linewidth]{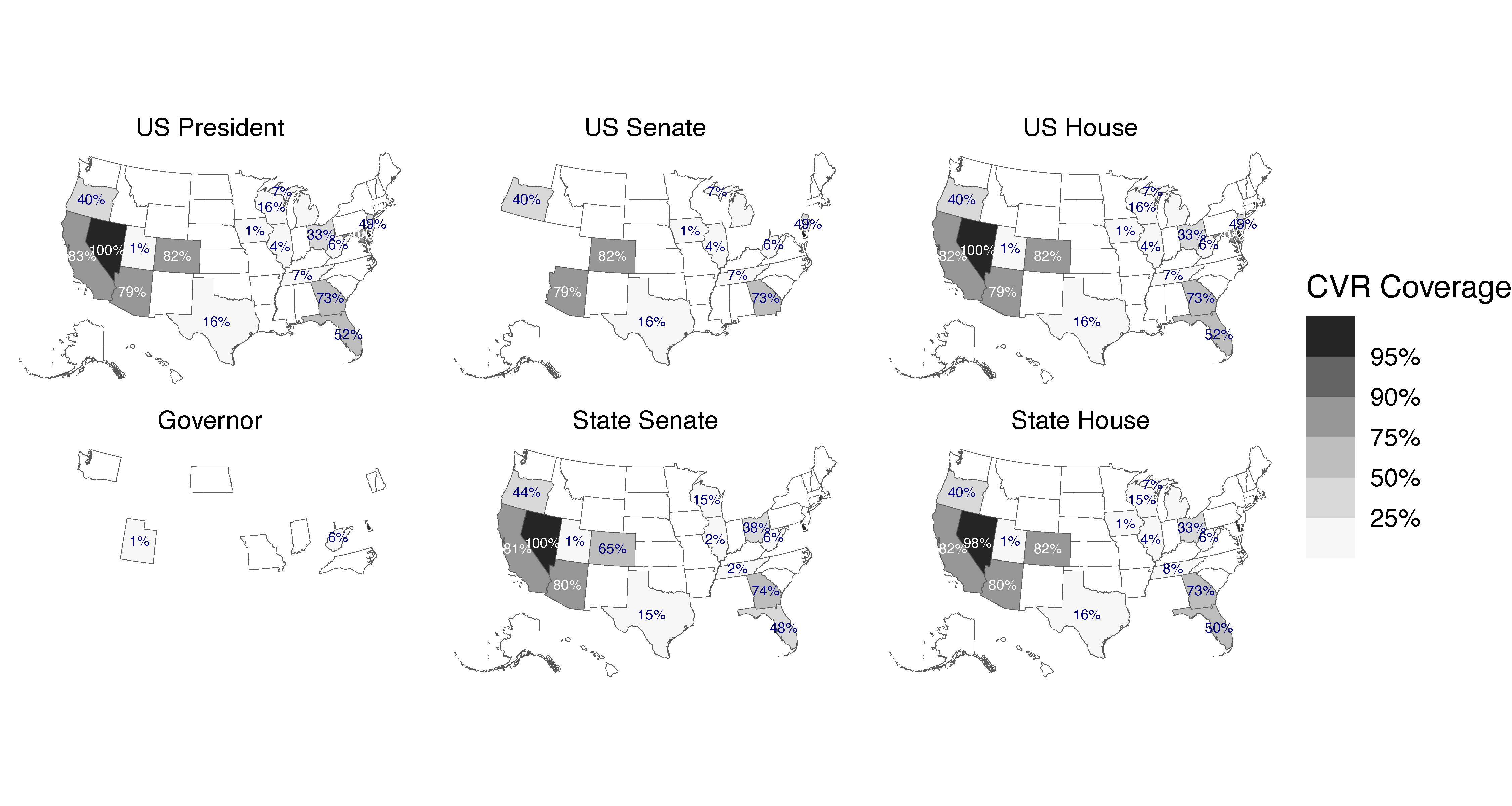}
\end{adjustwidth}
\end{figure}

\autoref{fig:map-frac} shows the states in our dataset and the coverage of its population for each office.
The percentage values assigned for each state indicate the total number of ballots relative to the total count of votes reported in the entire state.
For example, in Texas, we have around 15 percent of the votes for the President, Congress, and state legislature. There was no election for Governor in the state.

\newpage
\FloatBarrier

\section{Extracting Summaries from the Data}

In this section, we illustrate more examples of how users can extract useful summaries using our data. We open the dataset in the same manner as in the main text. 

Users should use the combination of \texttt{state}, \texttt{office}, and
\texttt{party} variable to identify candidates. The code below first
limits to vote choices for President in Wisconsin ballots using the
\texttt{filter()} command, and counts the number of records for each
candidate-party collection, sorted from most frequent to least.

\begin{Shaded}
\begin{Highlighting}[]
\NormalTok{ds }\SpecialCharTok{|\textgreater{}} 
  \FunctionTok{filter}\NormalTok{(state }\SpecialCharTok{==} \StringTok{"WISCONSIN"}\NormalTok{, office }\SpecialCharTok{==} \StringTok{"US PRESIDENT"}\NormalTok{) }\SpecialCharTok{|\textgreater{}} 
  \FunctionTok{count}\NormalTok{(candidate, party, }\AttributeTok{sort =} \ConstantTok{TRUE}\NormalTok{) }\SpecialCharTok{|\textgreater{}} 
  \FunctionTok{collect}\NormalTok{()}
\end{Highlighting}
\end{Shaded}

\begin{verbatim}
# A tibble: 8 x 3
  candidate       party      n
  <chr>           <chr>  <int>
1 DONALD J TRUMP  REP   586566
2 JOSEPH R BIDEN  DEM   442712
3 JO JORGENSEN    LBT    12544
4 WRITEIN         <NA>    2824
5 UNDERVOTE       <NA>    2674
6 BRIAN T CARROLL OTH     1748
7 DON BLANKENSHIP OTH     1448
8 OVERVOTE        <NA>     986
\end{verbatim}

For individual voters, use the \texttt{cvr\_id} variable within a state
and county. This ID is a numeric variable that is defined within
counties. These numbers do not in any way indicate the time in which the
ballot was cast, or the personal identity of the voter. The following
code extracts the vote from the voter marked with the \texttt{cvr\_id}
of 1.

\begin{Shaded}
\begin{Highlighting}[]
\NormalTok{ds }\SpecialCharTok{|\textgreater{}} 
  \FunctionTok{filter}\NormalTok{(state }\SpecialCharTok{==} \StringTok{"ARIZONA"}\NormalTok{, county\_name }\SpecialCharTok{==} \StringTok{"MARICOPA"}\NormalTok{) }\SpecialCharTok{|\textgreater{}} 
  \FunctionTok{filter}\NormalTok{(cvr\_id }\SpecialCharTok{==} \DecValTok{1}\NormalTok{) }\SpecialCharTok{|\textgreater{}} 
  \FunctionTok{select}\NormalTok{(county\_name, cvr\_id, office, district, candidate, party) }\SpecialCharTok{|\textgreater{}} 
  \FunctionTok{collect}\NormalTok{()}
\end{Highlighting}
\end{Shaded}

\begin{verbatim}
# A tibble: 6 x 6
  county_name cvr_id office       district candidate        party
  <chr>        <int> <chr>        <chr>    <chr>            <chr>
1 MARICOPA         1 US PRESIDENT FEDERAL  JOSEPH R BIDEN   DEM  
2 MARICOPA         1 US SENATE    ARIZONA  KELLY MARK       DEM  
3 MARICOPA         1 US HOUSE     008      MUSCATO MICHAEL  DEM  
4 MARICOPA         1 STATE SENATE 013      KERR SINE        REP  
5 MARICOPA         1 STATE HOUSE  013      DUNN TIMOTHY TIM REP  
6 MARICOPA         1 STATE HOUSE  013      SANDOVAL MARIANA DEM  
\end{verbatim}

This example shows that this voter split their ticket, voting for
Democrats in the Presidential and Congressional race, while voting for
one Republican candidate in State Senate. However, further investigation
into this voter's State Senate district shows that it was uncontested.
That is, with the following query,

\begin{Shaded}
\begin{Highlighting}[]
\NormalTok{ds }\SpecialCharTok{|\textgreater{}} 
  \FunctionTok{filter}\NormalTok{(state }\SpecialCharTok{==} \StringTok{"ARIZONA"}\NormalTok{, office }\SpecialCharTok{==} \StringTok{"STATE SENATE"}\NormalTok{, district }\SpecialCharTok{==} \StringTok{"013"}\NormalTok{) }\SpecialCharTok{|\textgreater{}} 
  \FunctionTok{count}\NormalTok{(candidate, party\_detailed) }\SpecialCharTok{|\textgreater{}} 
  \FunctionTok{collect}\NormalTok{()}
\end{Highlighting}
\end{Shaded}

\begin{verbatim}
# A tibble: 7 x 3
  candidate       party_detailed     n
  <chr>           <chr>          <int>
1 "KERR SINE"     REPUBLICAN     93388
2 "NOT QUALIFIED" <NA>            1852
3 "BACKUS BRENT"  <NA>             145
4 "UNDERVOTE"     <NA>           34391
5 ""              <NA>             119
6 "WRITEIN"       WRITEIN          531
7 "OVERVOTE"      <NA>              17
\end{verbatim}

\noindent We see that none of the ballots in State Senate district 13
were for a Democrat candidate, indicating that no Democrat ran in this
district. 
\end{document}